\documentclass[conference]{IEEEtran}%% Packages / general settings
\usepackage{graphicx}
\usepackage{amsmath,amssymb,dsfont,bbm,epstopdf,pgfplots} 
\usepackage{color}
\usepackage{cite}%hyperref}
\usepackage{graphics}
\usepackage{tikz}
\usepackage{pgfplots}
\usetikzlibrary{shapes, arrows, decorations.markings, arrows.meta}
\usetikzlibrary{patterns} % LATEX and plain TEX when using Tik Z
\allowdisplaybreaks
\graphicspath{{.}{./Figures/}} % path to figures
\allowdisplaybreaks[4] % Allows the align environment to wrap over pages.

\newtheorem{theorem}{Theorem}

\newtheorem{remark}{Remark}
\newtheorem{definition}{Definition}

\newcommand{\sK}{{\{1,\ldots,K\}}} 

\newcommand{\sRkf}{{\{1,\ldots,\lfloor 2^{n R_k^{(F)}} \rfloor \}}}
\newcommand{\sRks}{{\{1,\ldots,\lfloor 2^{n R_k^{(S)}} \rfloor \}}}
\newcommand{\km}{\mathsf{D}_{\max}}%{{\kappa_{\text{max}}}}

\renewcommand{\S}{{\sf{S}}}

\newcommand{\MkF}{M_k^{(F)}}
\newcommand{\MkS}{M_k^{(S)}}

\newcommand{\capa}{\mathsf{C}}

\definecolor{ForestGreen}{rgb}{0.0, 0.5, 0.0}

\begin{document}
\title{Mixed Delay Constraints in Wyner's Soft-Handoff Network}
\author{\IEEEauthorblockN{Homa Nikbakht and Mich\`ele Wigger}
	\IEEEauthorblockA{ LTCI, T$\acute{\mbox{e}}$l$\acute{\mbox{e}}$com ParisTech\\ %\\Universite Paris-Saclay,\\
		75013 Paris, France\\
		\{homa.nikbakht, michele.wigger\}@telecom-paristech.fr}
	\and
	\IEEEauthorblockN{Shlomo Shamai (Shitz)}
	\IEEEauthorblockA{Technion\\
		%Universite Paris-Saclay,\\
		sshlomo@ee.technion.ac.il}}
\maketitle

\begin{abstract}
	Wyner's \textit{soft-handoff} network with mixed delay constraints  is considered when neighbouring receivers can  cooperate over rate-limited  links. Each source message is a combination of independent ``fast'' and ``slow'' bits, where the former are subject to a stringent decoding delay. Inner and outer bounds on the capacity region are derived, and the  multiplexing gain region is characterized when only transmitters or only receivers  cooperate.
\end{abstract}

\section{Introduction}
Wireless communication networks have to accommodate  different types of data traffics with different  latency constraints. In particular, delay-sensitive video-applications represent an increasing portion of data traffic. On the other hand, modern networks can increase data rates by means of cooperation between terminals or with helper relays. However,  cooperation  typically introduces additional communication delays, and is thus not applicable to delay-sensitive applications.  In this paper, we analyze the rates of communication that can be attained over an interference network with either transmitter or receiver cooperation, and where parts of the messages cannot profit from this cooperation because they are subject to stringent delay constraints.  Mixed delay constraints in wireless networks have previously been studied in \cite{shlomo2012ISIT,Zhang2008,Zhang2005}. In particular,  \cite{shlomo2012ISIT} proposes a broadcasting approach over a single-antenna fading channel to   communicate a stream of ``fast" messages, which have to be  sent over a single coherence block, and a stream of ``slow" messages, which can be sent over   multiple blocks. %Dynamica resou
%\cite{Zhang2008} examined an optimal approach for dynamic resource allocation when both delay and Non-delay constrained streams are transmitted simultaneously. 
A similar approach was taken in  \cite{Zhang2005} but for a broadcast scenario with $K$ users. Instead of  superposing ``slow" on ``fast" messages, this latter work proposes a scheduling approach to give preference to  the communication of ``fast" messages.   The closely related setup of \emph{unreliable conferencing}, where a part of the message needs to be decoded without using the conferencing link,  was introduced in \cite{Steinberg, Steinberg2}. %The delay constraints are imposed to the applications anticipated for technologies such as 5G to address the latency issues. At the same time, these applications can experience a significant capacity increase by benefiting from the new paradigms, such as cooperative communications, that might delay the transmission. %One of the possible reasons that cooperation between users delays the communication is that the availability of cooperation links cannot be guaranteed due to the dynamic nature of the network \cite{Steinberg}.
%At the same time, these applications can experience a significant capacity increase by benefiting from new paradigms such as cooperative communications while  delay the transmission. 

For simplicity, in this paper, we focus on  Wyner's \textit{soft-handoff} model \cite{Wyner-94,Hanly-Whiting-93,Simeone-2011-Tut} with $K$ interfering transmitter and receiver pairs. Each transmitter sends a pair of independent source messages called ``fast'' and ``slow'' messages. Each receiver decodes the ``fast'' message immediately and only based on its own channel outputs. Before decoding its ``slow'' message, it  can communicate with its immediate neighbours over conferencing links during a given maximum  number of rounds \cite{Wiggeretal} and subject to a rate-constraint. It then decodes the ``slow" message based on its own channel outputs and the cooperation messages received from its neighbours. In the case of only transmitter conferencing, receivers decode both messages only based on their own channel outputs; transmitters can hold a conferencing communication that depends only on the ``slow" messages but not on the ``fast" messages.

We propose  inner and outer bounds on the capacity region of the soft-handoff network with receiver conferencing. We also  characterize the  multiplexing gain region of the setups with only transmitter conferencing or only receiver conferencing. The  multiplexing gain regions of the two scenarios  coincide, and thus show a   duality  between transmitter and receiver conferencing in the high signal-to-noise ratio regime. 

Our results also indicate that the sum-rate of ``fast" and ``slow" messages is approximately constant when ``fast" messages are sent at small  rate. In this regime,  the stringent decoding delay of part of the messages does not cause a loss in overall performance. When ``fast" messages have large rates, this is not the case. In this regime, increasing the rate of ``fast" messages by $\Delta$, requires that the rate of ``slow" messages be reduced by approximately  $2 \cdot \Delta$.  
%{\color{red}\par This paper is organized as follows, the problem setup is described in Section \ref{sec:DescriptionOfTheProblem}. Section \ref{Main Results} provides the main achieved results. Proofs of achievability and converse parts of the capacity region are derived in Section \ref{sec:scheme} and Section \ref{sec:converse}, respectively. }

\section{Problem Setup}\label{sec:DescriptionOfTheProblem}

%%%
%%%
%%%

{\color{black}Consider a wireless communication system as in  Fig. \ref{fig1} with $K$ interfering transmitter (Tx) and receiver (Rx) pairs  $1,\ldots, K$ that are aligned on a line. 
	 Transmitters and receivers are each equipped with a single antenna, and channel inputs and outputs are real valued. Interference is  short-range so that the signal sent by Tx~$(k)$ is observed only by Rx~$(k)$ and Rx~$(k+1)$. As a result, the time-$t$ channel output at Rx~$k$ is 
%We imagine a network with short-range interference, so that the signal sent by Tx~$k$ is only observed by receivers~$k$ and $k+1$. Specifically, the time-$t$ channel output at Rx~$k$ is 
\begin{equation}\label{Eqn:Channel}
Y_{k,t} = X_{k,t} + \alpha X_{k-1,t} +Z_{k,t},
\end{equation}
where $X_{k,t}$ and $X_{k-1,t}$ are the symbols sent by Tx~$(k)$ and $(k-1)$ at time $t$, respectively; $\{Z_{k,t}\}$ are independent and identically distributed (i.i.d.) standard Gaussians for all $k$ and $t$; $\alpha\neq 0$ is a fixed real number smaller than 1; and $X_{0,t} = 0$ for all $t$.}

Each Tx~$(k)$ wishes to send a pair of independent source messages  $M_k^{(F)}$ and $M_k^{(S)}$ to Rx~$(k)$. The ``fast" source message $\MkF$ is uniformly distributed over the set $
\mathcal{M}_{k}^{(F)}:=\sRkf$ and needs to be decoded subject to a  stringent delay constraint, as we explained shortly.  The ``slow" source message $\MkS$ is uniformly distributed over $\mathcal{M}_{k}^{(S)}:=\sRks$ and is subject to a less stringent decoding delay constraint.  Here, $n$ denotes the blocklength of transmission and $R_k^{(F)}$ and $R_k^{(S)}$ are the rates of transmissions of the ``fast" and the ``slow" messages.  
\begin{figure}[t]
%\small
  \centering
  \small
    %\hspace*{32pt}
 \begin{tikzpicture}[scale=1.47, >=stealth]
\centering
\tikzstyle{every node}=[draw,shape=circle, node distance=1cm];
\draw (-0.5,0) -- (0.5,0) -- (0.5,0.5) -- (-0.5,0.5) -- (-0.5,0);
\draw (-0.5,2) -- (0.5,2) -- (0.5,2.5) -- (-0.5,2.5) -- (-0.5,2);
\draw (1.5,2) -- (2.5,2) -- (2.5,2.5) -- (1.5,2.5) -- (1.5,2);
\draw (1.5,0) -- (2.5,0) -- (2.5,0.5) -- (1.5,0.5) -- (1.5,0);
\draw (3.5,2) -- (4.5,2) -- (4.5,2.5) -- (3.5,2.5) -- (3.5,2);
\draw (3.5,0) -- (4.5,0) -- (4.5,0.5) -- (3.5,0.5) -- (3.5,0);
\node [draw] at (0,1.25) {$+$};
\node [draw] at (2,1.25) {$+$};
\node [draw] at (4,1.25) {$+$};
\draw   [thick,->] (0,2)--(0,1.49);
\draw   [thick,->] (2,2)--(2,1.49);
\draw   [thick,->] (4,2)--(4,1.49);
\draw   [thick,->] (0,1.02)--(0,0.5);
\draw   [thick,->] (2,1.02)--(2,0.5);
\draw   [thick,->] (4,1.02)--(4,0.5);
\draw   [thick,->] (0.75,1.25)--(0.23,1.25);
\draw   [thick,->] (2.75,1.25)--(2.23,1.25);
\draw   [thick,->] (4.75,1.25)--(4.23,1.25);
\draw   [thick,->,dashed] (0,2)--node [draw=none,  text width=3.0cm, shape=rectangle, pos=0.5, yshift=-0.2cm,xshift=1.2cm ,rotate = -19,midway, fill=none, node distance=1cm] {\small{$\alpha_k X_{k-1}^n$}} (1.76,1.25);
\draw   [thick,->,dashed] (2,2)--node [draw=none,  text width=3.0cm, shape=rectangle, pos=0.5, yshift=-0.15cm,xshift=1.05cm ,rotate = -19,midway, fill=none, node distance=1cm] {\small{$\alpha_{k +1} X_{k}^n$}}(3.76,1.25);
\draw   [thick,->,dashed] (-1,1.55)--(-0.17,1.25);
\draw   [thick,->] (0.5,0.35)--(1.5,0.35);
\draw   [thick,->] (1.5,0.1)--(0.5,0.1);
\draw   [thick,->] (2.5,0.35)--(3.5,0.35);
\draw   [thick,->] (3.5,0.1)--(2.5,0.1);
\draw   [thick,->] (4.5,0.35)--(5,0.35);
\draw   [thick,->] (5,0.1)--(4.5,0.1);
\draw   [thick,->] (-1,0.35)--(-0.5,0.35);
\draw   [thick,->] (-0.5,0.1)--(-1,0.1);
%%%%%%%
\node [draw=none] at (-0.3,1.75) {\small{$X_{k-1}^n$}};
\node [draw=none] at (1.8,1.75) {\small$X_{k}^n$};
\node [draw=none] at (3.73,1.75) {\small$X_{k+1}^n$};
\node [draw=none] at (0.5,1.39) {\small$Z_{k-1}^n$};
\node [draw=none] at (2.5,1.39) {\small$Z_{k}^n$};
\node [draw=none] at (4.5,1.39) {\small$Z_{k+1}^n$};
\node [draw=none] at (-0.3,0.75) {\small$Y_{k-1}^n$};
\node [draw=none] at (1.8,0.75) {\small$Y_{k}^n$};
\node [draw=none] at (3.73,0.75) {\small$Y_{k+1}^n$};
\node [draw=none] at (0,0.25) {\small Rx $k-1$};
\node [draw=none] at (2,0.25) {\small Rx $k$};
\node [draw=none] at (4,0.25) {\small Rx $k+1$};
\node [draw=none] at (0,2.25) {\small Tx $k-1$};
\node [draw=none] at (2,2.25) {\small Tx $k$};
\node [draw=none] at (4,2.25) {\small Tx $k+1$};
%\node [draw=none] at (0,2.65) {\small{$\small{M_{k-1} : \{M_{k-1}^{(F)},M_{k-1}^{(S)}\}}$}};
%\node [draw=none] at (2,2.65) {\small{$\small{M_{k} : \{M_{k}^{(F)},M_{k}^{(S)}\}}$}};
%\node [draw=none] at (4,2.65) {\small{$\small{M_{k+1} : \{M_{k+1}^{(F)},M_{k+1}^{(S)}\}}$}};
\node [draw=none] at (1,0.55) {\small{$Q^{(j)}_{k-1 \rightarrow k}$}};
\node [draw=none] at (1,-0.1) {\small{$Q^{(j)}_{k \rightarrow k-1}$}};
\node [draw=none] at (3,0.55) {\small{$Q^{(j)}_{k \rightarrow k+1}$}};
\node [draw=none] at (3,-0.1) {\small{$Q^{(j)}_{k+1 \rightarrow k}$}};
%\node [draw=none] at (3,0.4) {$\hat{{M}}_{k}^{(F)}$};
%\node [draw=none, rotate = 90] (v1) at (1.58,0.4) {$\ldots$};
%\draw   [thick,latex'-latex'] (1.5,1.1)-- node [draw=none,  text width=3.0cm, shape=rectangle, pos=0.5, yshift=0.2cm,xshift=0.6cm ,midway, fill=none, node distance=1cm] {\small{$K$\text{subfiles}}}  (3,1.1);
%\draw   [thick,latex'-latex'] (1.4,1)-- node [draw=none,  text width=1.7cm, shape=rectangle, pos=0.5, yshift=3,xshift=-5 ,midway, fill=none, node distance=1cm,rotate = 90] {\small{$m\;$\text{files}}}  (1.4,0);
%\node [draw=none] at (2.25,-0.2) {$\tilde{L}$};
\end{tikzpicture}
\vspace*{-5ex}

  \caption{System model}
  \label{fig1}
  \vspace*{-2ex}
\end{figure}
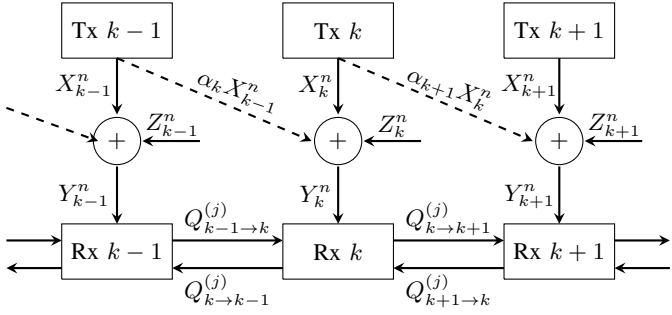~~%
%%%%%%%%%%%%%%%%%%%%%%%%%%%%%%

%%%%%%%%%%%%%%%%%%%%%%%%%%%%%%%%%%%%%%%%%%%%

%\begin{figure}
%\begin{center}
%\includegraphics[width=\columnwidth]{AsymmetricWynerNetwork}
All source messages are  independent of each other and of all channel noises. 

Tx~$(k)$ computes its channel inputs $X_k^n:=(X_{k,1},\ldots, X_{k,n})$ as a function of the pair $(M_{k}^{(F)}, M_{k}^{(S)})$:
\begin{equation}
X_{k}^n = f_k^{(n)} \big( M_{k}^{(F)}, M_{k}^{(S)}\big),
\end{equation} for some function $f_k^{(n)}$ on appropriate domains that satisfies % \colon \mathcal{M}_k^{(F)} \times \mathcal{M}_k^{(S)} \rightarrow \mathbb{R}^n.
 the average block-power constraint
\begin{equation}\label{eq:power}
\frac{1}{n} \sum_{t=1}^n X_{k,t}^2
\leq P, \quad \text{a.s.},
\quad \forall\ k \in \sK.
\end{equation}

Receivers  decode in two phases. During the first \emph{fast-decoding phase}, each  Rx~$(k)$ decodes the ``fast" source message $M_k^{(F)}$ based on its own channel outputs $Y_k^n:=(Y_{k,1},\ldots, Y_{k,n})$. So, it produces:
\begin{align}
	&\hat{{{M}}} _k^{(F)} ={g_k^{(n)}}\big( Y_k^{n}\big)
	\end{align} 
for some decoding function $g_k^{(n)}$ on appropriate domains.
	
	In the subsequent \emph{slow-decoding phase}, the receivers first communicate to each other over orthogonal conferencing links, and then they decode their intended ``slow" messages based on their own channel outputs and the conferencing messages received from their neighbours.  Only neighbouring receivers can exchange conferencing messages, and conferencing  is limited to a maximum number of $\km$  rounds and to a rate-constraint $\pi$. In   conferencing round $j \in \{1,2, \dots, \km\}$, Rx~$(k)$ sends the conferencing message $Q^{(j)}_{k\rightarrow k-1}$ to its left neighbour, Rx~$(k-1)$, and the conferencing message $Q^{(j)}_{k\rightarrow k+1}$ to its right neighbour, Rx~$(k+1)$. These conferencing messages can depend on the outputs $Y_k^n$  and on the conferencing messages that Rx~$(k)$ received in the previous rounds. So, for $\tilde{k} \in \{k-1,k+1\}$:
	\begin{align}
	&Q^{(j)}_{k\rightarrow \tilde k}={\psi_{k,\tilde k}^{(n)}}\Big( Y_k^n, Q^{(1)}_{k-1\rightarrow k},Q^{(1)}_{k+1\rightarrow k},\ldots , \nonumber \\
	 & \hspace{5cm}Q^{(j-1)}_{k-1\rightarrow k},Q^{(j-1)}_{k+1\rightarrow k} \Big),  
	\end{align}
	for an   encoding function $\psi_{k,\tilde k}^{(n)}$  on appropriate domains. The $\km$ messages sent over a  conferencing link in each direction are  subject to a rate constraint $\pi$. So, for all $k\in\{1,\ldots, K\}$  and $\tilde{k}\in\{k, k+1\}$:   
	\begin{equation}\label{eq:conference_capa}
\sum_{j=1}^{\km}	H( Q^{(j)}_{k\rightarrow \tilde k})	 \leq  \pi \cdot n.
		\end{equation}
%	\begin{align}
%	&\psi_{k,\tilde k}^{(n)}\colon {\mathcal{M}}_k^{(F)} \times \prod_{l=1}^{j-1}\prod _{\tilde k \in \{k-1,k+1\}}\sCl \rightarrow  \sCl.
%	\end{align} 

After the last conferencing round $\km$, each Rx~$(k)$ decodes its desired ``slow" message as
	\begin{align}
	&\hat{M}_k^{(S)}:={b_k^{(n)}}\Big( Y_k^n, Q^{(1)}_{k-1\rightarrow k},Q^{(1)}_{k+1\rightarrow k}, \notag \\& \hspace{4.5cm}\ldots, Q^{(\km)}_{k-1\rightarrow k},Q^{(\km)}_{k+1\rightarrow k} \Big) 
	\end{align}
	by means of a decoding function $b_{k}^{(n)} $ on appropriate domains. %:\hat{\mathcal{Y}}_k^{n} \times \prod_{l=1}^{\km}\prod _{\tilde k \in \{k-1,k+1\}}\rightarrow \mathcal{M}_k$.

The main interest in this paper is in the  achievable sum-rates of ``fast" and ``slow" messages. Given a maximum conferencing rate $\pi$ and a maximum allowed power  $P$, the pair of (average) rates $(R^{(F)}, R^{(S)})$ is called \emph{achievable}, if there exists a sequence (in $n$) of encoding and decoding functions so that 
\begin{equation}
\frac{1}{K} \sum_{k=1}^K R_{k}^{(F)}  = R^{(F)} \qquad \textnormal{and} \qquad \frac{1}{K} \sum_{k=1}^K  R_{k}^{(S)}  = R^{(S)},
\end{equation}
and the probability of decoding error 
 \begin{align*}
P_e^{(n)}:= \text{Pr}\bigg[\bigcup_{k\in\{1,\ldots, K\}} \Big \{ \hat{{M}}_k^{(F)} \neq {M}_k^{(F)} \textnormal{ or } \hat{{M}}_k^{(S)} \neq {M}_k^{(S)}\Big\} \bigg]
 \end{align*}
 tends to 0 as $n\to \infty$.
 
 \begin{definition}
 	Given power constraint $P>0$ and maximum conferencing rate $\pi$, the \emph{capacity region} $\capa(P,\pi)$ is the closure of the set of all rate pairs $(R^{(F)}, R^{(S)})$ that are achievable.
 	\end{definition}
 	
 	We will particularly be interested  in the high signal-to-noise ratio (SNR) regime, and thus  in the set of achievable \emph{multiplexing gains} when the conferencing capacity also scales logarithmically in the  SNR. Given a conferencing prelog $\mu\geq 0$, the \emph{pair of  multiplexing gains} $(\S^{(F)},\S^{(S)})$ is called \emph{achievable}, if for each $K$ there exists a sequence of  rates $\{R_K^{(F)}(P),R_K^{(S)}(P) \}_{P>0}$  so that 
  	 	\begin{align} \label{sf}
  	 	\S^{(F)}&:= \varlimsup_{K\rightarrow \infty}\; \varlimsup_{P\rightarrow\infty} \;\frac{ R_K^{(F)}}{\frac{1}{2}\log (1+P)},\\\label{ss}
  	 	\S^{(S)}&:= \varlimsup_{K\rightarrow \infty}\; \varlimsup_{P\rightarrow\infty} \;\frac{ R_K^{(S)}}{\frac{1}{2}\log (1+P)},
  	 	\end{align}
and for each $K$ and  $P>0$ the pair $(R_K^{(F)}(P),R_K^{(S)}(P))$ is achievable with conferencing rate at most $\pi= \mu \cdot \frac{1}{2} \log P$.
 	
 	 %%%%%%%%%%%%%%
 	 \begin{definition}
 	 Given a conferencing-prelog $\mu$, the  closure of the set of all achievable   multiplexing gains $(\S^{(F)}, \S^{(S)})$ is called \emph{multiplexing gain region} and denoted $\mathcal{S}^\star(\mu)$.
 	 \end{definition}
 \section{Main Results} \label{Main Results}
 
 Our first result is an inner bound on the capacity region. It is based on two schemes. The first scheme assumes $\pi \le R^{(F)}$.  Each transmitter uses a 3-layer superposition code, where it sends its  ``fast" message in the lowest two layers and its ``slow" message in the upper-most layer. Each receiver immediately decodes its intended ``fast"   message based only on its channel outputs and then sends the part encoded in the lower-most layer to its right neighbour. To decode its intended ``slow" message, it first pre-subtracts the interference caused by the lower-most layer of the superposition codeword sent by the transmitter to its left. This scheme uses only a single conferencing round. 
 
 The second scheme assumes $\pi > R^{(F)}$ and also exchanges parts of ``slow" source messages over the conferencing links. Each transmitter employs a $\km+1$-layer superposition code, where the lower-most layer encodes the ``fast" message and all higher layers encode parts of the ``slow" message. As before, each receiver decodes its intended ``fast" message  immediately based on its channel outputs. It then sends this  decoded message  over the conferencing link to its left neighbour during the first conferencing round. Subsequently, after each conferencing round $j=1,\ldots, \km$, each receiver cancels  the interference from the layer-$j$ codeword sent by the transmitter to its left and then decodes the layer-$j+1$ of its intended message. It sends the decoded  message part and the conferencing message that it obtained in the previous rounds to its right neighbour. 
 
\begin{theorem}[Capacity Inner Bound] \label{lemma1} 
The capacity region $\capa(P, \pi)$ includes all  rate-pairs $(R^{(F)}, R^{(S)})$ that   satisfy 
\begin{subequations}\label{eq:ach1}
 \begin{align} \label{rsf}
 R^{(F)} & \le  \min \big\{I(U_{2}; Y), I(U_{2}; Y| U_{1}) + \pi\big\}  
 \end{align}and
 \begin{align}
 R^{(F)} + R^{(S)} & \le \frac{1}{K} \sum_{k=1}^K \Big[  I(X; Y, U_{1}'| U_{1}) \nonumber \\
 & \qquad\quad + \min \big\{I(U_{2}; Y), I(U_{2}; Y| U_{1}) + \pi\big\}\Big],
 \end{align}
 \end{subequations}
 where triples $(U_{1}, U_{2}, X)$ and $(U_{1}', U_{2}', X')$ are i.i.d. 
 according to some probability distribution $P_{U_1U_2X}$ that satisfies the Markov chain $U_1 \to U_2 \to X$, and where $Y= X +\alpha X' +Z$ with $Z$ standard Gaussian independent of $(U_1,U_2,X,U_1', U_2',X')$.

The capacity region $\capa(P, \pi)$ also  includes all  rate-pairs $(R^{(F)}, R^{(S)})$ that   satisfy 
\begin{subequations}\label{eq:ach2}
  \begin{align} \label{rsf2}
 R^{(F)} & \le   I(U; Y) \\
R^{(F)}+  R^{(S)} & \le I(U;Y) + I(V_{1};Y, U'| U )  \nonumber \\
 & \qquad \; + \sum_{d = 2}^{\km-1} I(V_{d}; Y, V_{d-1}' | V_{d-1})\nonumber \\
 & \qquad  +I(X;Y,X'|V_{\km-1}),
 \end{align}
 \end{subequations}
  where the tuples $(U, V_{1}, \ldots, V_{\km-1},X)$ and $(U', V_{1}', \ldots, V_{\km-1}',X')$ are i.i.d. according to some probability distribution $P_{UV_{1}\ldots V_{\km-1}X}$ satisfying the Markov chain $U\to V_1 \to V_2 \to \ldots \to V_{\km-1}\to X$ and the rate constraint % $P_{UV_1}\cdot P_{V_2|V_1}\cdot$ $ P_{V_3|V_2}\cdots P_{V_{\km-1} |V_{\km-2}}$ and satisfies
  \begin{equation}
  I(U; Y)+ I(V_{1};Y, U'| U ) 
+ \sum_{d = 2}^{\km-1} I(V_{d}; Y, V_{d-1}' | V_{d-1}) \leq \pi, 
  \end{equation}
  and where $Y= X +\alpha X' +Z$ with $Z$ independent standard Gaussian.
 \end{theorem}
 \begin{IEEEproof}
 	See Section~\ref{sec:scheme}.
 	\end{IEEEproof}

 %%%%%%%%%%%%%%%%%%%%%%%%%%
  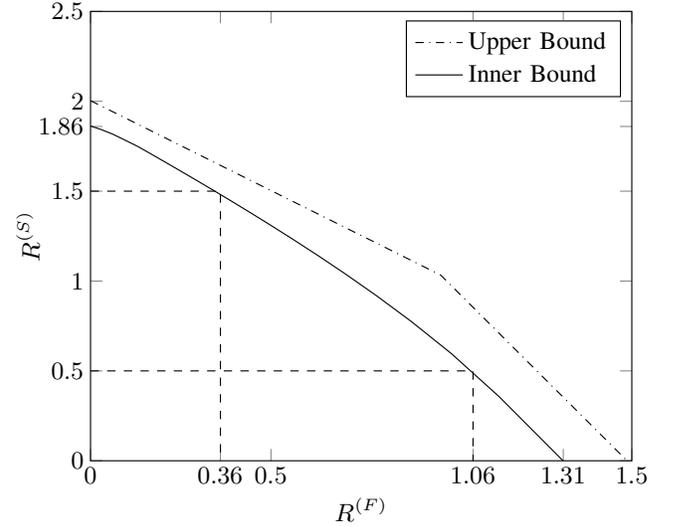
\begin{figure}[!t]
\centering
\begin{tikzpicture}[scale=1.05,font=\tiny]
\begin{axis}[
    xlabel={\small {$R^{(F)}$ }},
    ylabel={\small {$R^{(S)}$ }},
     xlabel style={yshift=.5em},
     ylabel style={yshift=-1.25em},
    xmin=0, xmax=1.5,
    ymin=0, ymax=2.5,
    xtick={0,0.36,0.5,1.06,1.31,1.5},
    ytick={0,0.5,1,1.5,1.86,2,2.5},
    yticklabel style = {font=\small,xshift=0.25ex},
    xticklabel style = {font=\small,yshift=0.25ex},
    legend style = {nodes = right}
]

\addplot[
    color=black,
      dash dot
    ]
    coordinates {
    (0,2.0035)(0.9702,1.0333)(1.4868,0)
    };%conventional t
    
    \addplot[
    color=black,
    ]
    coordinates {
 (0,1.8627896006476)( 0.0307002723320716,1.84201326922935)(0.0627654410419294,1.81603834828985)(0.0963225389711979,1.78317468520822)(0.131517202916897,1.74821291305975)(0.168517493638785,1.70590679920228)	(0.207518749639422,1.66096404744368)(0.248749829735408,1.61303403973992)(0.292481250360578,1.56169120775264)(0.339035952556319,1.50641202017879)(0.388803789331776,1.44654239804174)	(0.442261391290032,1.38125034313667) (0.500000000000000,1.30945491632225)(0.562765441041929,1.22971580931865)(0.631517202916897,1.14005395959637)(0.707518749639422,1.03764406365212)	(0.792481250360578,0.918250633858560)(0.888803789331776,0.775098541280240)(1,0.596322538971198)(1.13151720291690,0.358103516999704)(1.30948125036058,0)
    };%conventional 
   \legend{{\small Upper Bound},{\small Inner Bound}}  
   \addplot[
    color=black,
      dashed
    ]
    coordinates {
    (0,1.5)(0.36,1.5)(0.36,0)
    };%conventional t
    
     \addplot[
    color=black,
      dashed
    ]
    coordinates {
    (0,0.5)(1.06,0.5)(1.06,0)
    };%conventional t
\end{axis}
\end{tikzpicture}
\vspace{-0.25cm}
\caption{Capacity outer bound in Theorem \ref{thm:converse} and inner bound in Theorem \ref{lemma1}  for $P = 5$, $\alpha = 0.2$, $\pi=0.346$, and $\km=16$.}
\label{fig4}
%\vspace{-0.5cm}
\end{figure}
  \begin{figure}[!t]
\centering
\begin{tikzpicture}[scale=1.05,font=\tiny]
\begin{axis}[
    xlabel={\small {$R^{(F)}$ }},
    ylabel={\small {$R^{(S)}$ }},
     xlabel style={yshift=.5em},
     ylabel style={yshift=-1.25em},
    xmin=0, xmax=1.4,
    ymin=0, ymax=3.5,
    xtick={0,0.2,0.4,0.6,0.8,1,1.2,1.4},
    ytick={0,0.5,1,1.5,2,2.5,3,3.5},
    yticklabel style = {font=\small,xshift=0.25ex},
    xticklabel style = {font=\small,yshift=0.25ex},
    legend style = {nodes = right}
]

     \addplot[
    color=black,dash dot,
    ]
    coordinates {
(0,2.86844951790305)	(0.0307002723320716,2.83216076508266)	(0.0627654410419294,2.79394928445036)	(0.0963225389711979,2.75359987240706)	(0.131517202916897,2.71085904195156)	(0.168517493638785,2.66542536658497)	(0.207518749639422,2.61693657369844)	(0.248749829735408,2.56495197899321)	(0.292481250360578,2.50892808568201)	(0.339035952556319,2.44818388536702)	(0.388803789331776,2.38185016008902)	(0.442261391290032,2.30879302611250)	(0.500000000000000,2.22749422027644)	(0.562765441041929,2.13585496837824)	(0.631517202916897,2.03085621457981)	(0.707518749639422,1.90792696090319)	(0.792481250360578,1.75965612221981)	(0.888803789331776,1.57280864478984)	(1,1.31997345806964)	(1.13151720291690,0.927305780922377)	(1.30948125036058,0)
    };%conventional 
 
     \addplot[
    color=black, dashed
    ]
    coordinates {
(0,2.73366218021606)	(0.0307002723320716,2.69751952768778)	(0.0627654410419294,2.65947000861807)	(0.0963225389711979,2.61930115061908)	(0.131517202916897,2.57676286314920)	(0.168517493638785,2.53155799709676)	(0.207518749639422,2.48332974357539)	(0.248749829735408,2.43164451027606)	(0.292481250360578,2.37596817605815)	(0.339035952556319,2.31563239149182)	(0.388803789331776,2.24978545608025)	(0.442261391290032,2.17731844029539)	(0.500000000000000,2.09674988697383)	(0.562765441041929,2.00603766391255)	(0.631517202916897,1.90225463712127)	(0.707518749639422,1.78098954475542)	(0.792481250360578,1.63513553293625)	(0.888803789331776,1.45211741618554)	(1,1.20627393227393)	(1.13151720291690,0.830434381976273)	(1.30948125036058,0)
    };%conventional 
     \addplot[
    color=black, dotted
    ]
    coordinates {
(0,2.56396778608285)	(0.0307002723320716,2.52805242576934)	(0.0627654410419294,2.49025473728091)	(0.0963225389711979,2.45036644871488)	(0.131517202916897,2.40814268675527)	(0.168517493638785,2.36329286370459)	(0.207518749639422,2.31546853898874)	(0.248749829735408,2.26424696348588)	(0.292481250360578,2.20910832208170)	(0.339035952556319,2.14940353651730)	(0.388803789331776,2.08430750134155)	(0.442261391290032,2.01274905050710)	(0.500000000000000,1.93330220923000)	(0.562765441041929,1.84400982018891)	(0.631517202916897,1.74208181728468)	(0.707518749639422,1.62334324582542)	(0.792481250360578,1.48113246829643)	(0.888803789331776,1.30382481603765)	(1,1.06821589753806)	(1.13151720291690,0.716043103475406)	(1.30948125036058,0)
    };%conventional 
     \addplot[
    color=black,  
    ]
    coordinates {
(0,2.33635339216008)	(0.0307002723320716,2.30084012662052)	(0.0627654410419294,2.26348753350060)	(0.0963225389711979,2.22409462848322)	(0.131517202916897,2.18242557122923)	(0.168517493638785,2.13820110822918)	(0.207518749639422,2.09108721698182)	(0.248749829735408,2.04067977575706)	(0.292481250360578,1.98648346326439)	(0.339035952556319,1.92788206838716)	(0.388803789331776,1.86409563872133)	(0.442261391290032,1.79411677990322)	(0.500000000000000,1.71661260768952)	(0.562765441041929,1.62976741268817)	(0.631517202916897,1.53101702403717)	(0.707518749639422,1.41657091679420)	(0.792481250360578,1.28047911918309)	(0.888803789331776,1.11259832004821)	(1,0.893402194894693)	(1.13151720291690,0.576983309374080)	(1.30948125036058,0)
    };%conventional 
   \legend{{\small $\km = 10$},{\small $\km = 8$},{\small $\km = 6$},{\small $\km = 4$}}  

\end{axis}
\end{tikzpicture}
\vspace{-0.25cm}
\caption{Capacity inner bound in Theorem \ref{lemma1}  for $\pi = 2$, $P =5$, $\alpha = 0.2$ and different values of $\km$.}
\label{fig4-3}
%\vspace{-0.5cm}
\end{figure}
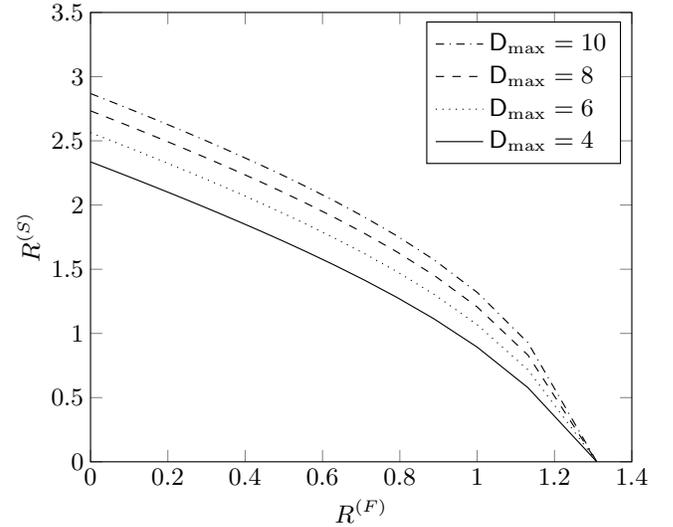
%%%%%%%%%%%%%%%%%%%%%%

 \begin{theorem}[Capacity Outer Bound]\label{thm:converse}
 	Any  achievable  rate pair $(R^{(F)}, R^{(S)})$ satisfies the following two conditions: 
 	\begin{IEEEeqnarray}{rCl}
 		R^{(F)} +R^{(S)}	& \leq &\frac{  \left( \left \lceil\frac{K-1}{2} \right \rceil+1\right) }{K}  \cdot  \frac{1}{2} \log \left(1+ (1+\alpha^2)P\right)  \nonumber \\[1ex]
 		& &\hspace{0cm}+  \frac{   \left \lfloor\frac{K-1}{2} \right \rfloor }{K} \cdot  \max \{-\log|\alpha|, 0 \}   \nonumber \\[1ex]
 		 & & +\frac{  \left \lfloor \frac{K}{2} \right \rfloor }{K}  \cdot  \frac{1}{2} \log (1+\alpha^2)+  \frac{K-1}{K} \cdot \pi   ,\label{eq:conv1}\\[1.2ex]
 2 R^{(F)} +R^{(S)}	& \leq &  \frac{K-1}{2K} \bigg( \frac{1}{2}\log \big((1+(1+ \alpha^2) P) (1+\alpha^2)\big) \nonumber \\
 && \hspace{3cm}+2 \max \{ -  \log |\alpha| ,0\} \bigg)\nonumber \\
  && +\frac{1}{K} \log( 1+P) .\label{eq:conv2}
 	\end{IEEEeqnarray}
 \end{theorem}
  \begin{IEEEproof}
  	See Section~\ref{sec:converse}.
  \end{IEEEproof}
  
  Fig.~\ref{fig4}  illustrates the outer bound on the capacity-region in Theorem~\ref{thm:converse} and the inner 
bound in Theorem~\ref{lemma1} when this latter is evaluated for jointly Gaussian distributions on the inputs and the auxiliaries. For small values of $R^{(F)}$, both the lower and the upper bounds decrease with slope -1. For large values of $R^{(F)}$, they decrease with slope -2.  % order to illustrate the capacity regions, in case of $\pi < R^{(F)}$, it is assumed that $U_{1}\sim \mathcal{N}(0,\beta)$, $U_{2}= U_{1} +   \mathcal{N}(0,\delta-\beta)$ and $X= U_{2} +   \mathcal{N}(0,1-\delta)$. For $\pi > R^{(F)}$, we consider $U \sim \mathcal{N}(0,\beta)$, $V_{1} = U + \mathcal{N}(0,\delta_1-\beta)$ and $V_{d}= V_{d-1} + \mathcal{N}(0,\delta_d-\delta_{d-1})$ with $d \in \{2,\ldots,\km-1\}$. The acheived results are shown in   Fig. \ref{fig2} for different values of $\pi$ and $\km$.   
% \begin{figure}[!t]
% 	%%\psfragscanon
% %	\centering
% 	%                     %\psfrag{Density of Nodes (\lambda)}[cl][cl][1.]{Density of Nodes ($\lambda$)}                   
% 	\includegraphics[width=1.1\linewidth]{RFRSf2.eps}
% 	\caption{Inner bound in Theorem~\ref{lemma1} for different values of $\pi$ and $\km$.}
% 	\label{fig2}
% \end{figure}
 
 %%%%%%%%%%%%%%
 
Inner and outer bounds  in Theorems~\ref{lemma1} and \ref{thm:converse} are  generally not tight at high SNR. New bounds are required to obtain the following theorem. In particular, a new inner bound based on a scheme that  periodically silences transmitters so as to avoid that interference propagates too far.
 
\begin{theorem}[Multiplexing Gain] \label{lemma3} 
The multiplexing gain  region $\mathcal{S}^\star(\mu)$ is the set of all nonnegative pairs $(\S^{(F)}, \S^{(S)})$  satisfying
 \begin{align} 
 2\S^{(F)} + \S^{(S)} & \le 1   \label{mg1}\\
 \S^{(F)} + \S^{(S)} & \le \min\bigg\{  \frac{1}{2} + {\mu},\;\; \frac{ 2 \km +1}{2 \km+2} \bigg\}.\label{mg2}
 \end{align}
 \end{theorem}
 
 \begin{IEEEproof}
 	See Section~\ref{sec:rxconf}.
 	\end{IEEEproof}
 
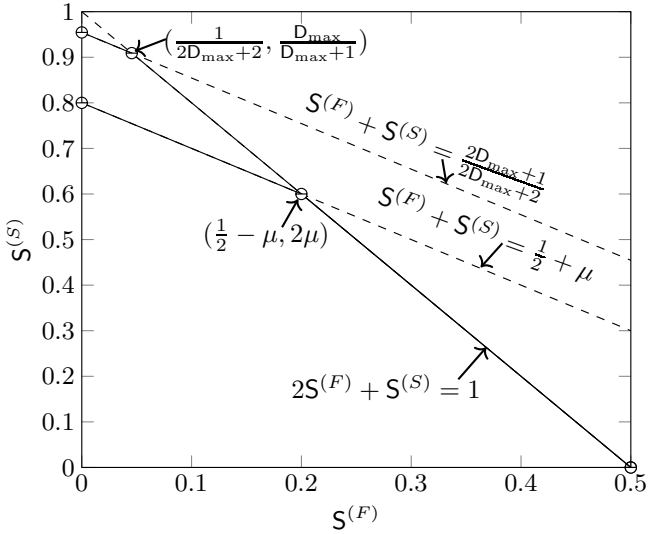
\begin{figure}[!t]
\centering
\begin{tikzpicture}[scale=1.065]
\begin{axis}[
    xlabel={\small {$\S^{(F)}$ }},
    ylabel={\small {$\S^{(S)}$ }},
     xlabel style={yshift=.5em},
     ylabel style={yshift=-1.25em},
    xmin=0, xmax=0.5,
    ymin=0, ymax=1,
    xtick={0,0.1,0.2,0.3,0.4,0.5},
    ytick={0,0.1,0.2,0.3,0.4,0.5,0.6,0.7,0.8,0.9,1},
    yticklabel style = {font=\small,xshift=0.25ex},
    xticklabel style = {font=\small,yshift=0.25ex},
]

\addplot[
    color=black,
    mark=halfcircle,
    ]
    coordinates {
    (0,0.8)(0.2,0.6)(0.5,0)
    };%1
\addplot[
    color=black,
    mark=halfcircle,
    ]
    coordinates {
    (0,0.9545)(0.0455, 0.9091)(0.5,0)
    };%2
\addplot[
    color=black,
      dashed
    ]
    coordinates {
    (0,1)(0.5,0)
    };%dashed1
    
    \addplot[
    color=black,
      dashed
    ]
    coordinates {
     (0,0.9545) (0.9545,0)
    };%dashed2
    
     \addplot[
    color=black,
      dashed
    ]
    coordinates {
     (0,0.8) (0.8,0)
    };%dashed3
    
\end{axis}
\draw [->,thick](2.5,3)--(2.7,3.35);
\node [draw=none] (v1) at (2.3,2.9) {$(\frac{1}{2}-\mu, 2\mu)$};
\draw [->,thick](4.7,1.1)--(5.05,1.5);
\node [draw=none] (v1) at (3.8,1) {$2\S^{(F)} + \S^{(S)} = 1$};
\draw [->,thick](5.25,2.8)--(4.98,2.5);
\node [draw=none, rotate = -20] (v1) at (5.05,2.9) {$\S^{(F)} + \S^{(S)} = \frac{1}{2} + \mu$};
\draw [->,thick](4.5,3.85)--(4.55,3.55);
\node [draw=none, rotate = -20] (v1) at (4.3,4) {$\S^{(F)} + \S^{(S)} = \frac{ 2 \km +1}{2 \km+2}$};
\draw [->,thick](1.05,5.35)--(0.66,5.2);
\node [draw=none] (v1) at (2.3,5.3) {$(\frac{1}{2\km+2}, \frac{\km}{\km+1})$};
\end{tikzpicture}
\vspace{-0.75cm}
\caption{Region $\mathcal{S}^\star(\mu)$ for $\km=10$ and $\mu= 0.3$ or $\mu\geq 10/22$.}
\label{fig3}
%\vspace{-0.5cm}
\end{figure}

 Figure~\ref{fig3} shows the  multiplexing gain region for different values of $\mu$. We notice that for $\S^{(F)} \leq  \frac{1}{2}-\mu$, the slope of the boundary of the region  is $-1$, and for $\S^{(F)} >  \frac{1}{2}-\mu$, it is $-2$.
  
% \begin{figure}[!t]
% 	%%\psfragscanon
% 	\centering
% 	%                     %\psfrag{Density of Nodes (\lambda)}[cl][cl][1.]{Density of Nodes ($\lambda$)}                   
% 	\includegraphics[width=1.1\linewidth]{SFSS2.eps}
% 	\caption{Multiplexing gain region $\mathcal{S}^\star(\mu)$ for $\km=10$ and $\mu= 0.3$ and $10/22$.}%\mu \geq \frac{\km}{2\km+2}$.}
% 	\label{fig3}
% \end{figure}
 
 %%%%%%%%%%%%%%%%%%%%%
\vspace{0cm}
 %%%%%%%%%%%%%%%%%%%%%%%%%%%%%%%%%%%%%
 
 \begin{remark}
An analogous result can be obtained for the  setup where each receiver can send conferencing messages only to its left neighbour or only to its right neighbour. The  multiplexing gain region is characterized by \eqref{mg1} and  
\begin{equation}
 \S^{(F)} + \S^{(S)} \leq \min\bigg\{  \frac{1}{2} + \frac{\mu}{2},\;\; \frac{  \km +1}{ \km+2} \bigg\}.
\end{equation}
Notice that despite the asymmetry of the network, the result is the same for conferencing to left   or  right neighbours.
 \end{remark}
 %%%%%%%%%%%%%%%%%%%%%%
 \section{Transmitter-Conferencing}
We consider a related setup where transmitters can send conferencing messages but not the receivers. Transmitter conferencing is limited to $\km$ rounds and the exchanged  messages  can only depend on the ``slow" messages but not on the fast messages. This models a setup where  the transmitters learn the ``slow" messages in advance before they communicate to the receivers, whereas ``fast" messages arrive at the transmitters just shortly before this communication.

In each round $j\in\{1,\ldots, \km\}$, Tx~$(k)$ produces  the two conferencing messages $T_{k\to k-1}^{(j)}$ and $T_{k\to k+1}^{(j)}$, where
\begin{IEEEeqnarray}{rCl}
T_{k\to \tilde{k}}^{(j)}  &= &\xi_{k\to \tilde{k}}^{(n)} \big(M_{k}^{(S)}, T_{k-1\to k}^{(1)}, \ldots, T_{k-1\to k}^{(j-1)},  \nonumber \\
 & & \hspace{3cm} T_{k+1\to k}^{(1)}, \ldots, T_{k+1\to k}^{(j-1)}\big) \IEEEeqnarraynumspace
\end{IEEEeqnarray}
for some function $\xi_{k \to \tilde{k}}^{(n)}$ on appropriate domains. It  sends these messages over the conferencing links to its left and right neighbours. As before, the conferencing links are rate-limited to rate $\pi$. So, for all $k\in\{1,\ldots, K\}$  and $\tilde{k}\in\{k-1, k+1\}$:  
        \begin{equation}\label{eq:conference_capa}
\sum_{j=1}^{\km}        H( T^{(j)}_{k\rightarrow \tilde k})         \leq  \pi \cdot n.
                \end{equation}
        Each Tx~$(k)$ then computes its channel inputs as
        \begin{IEEEeqnarray}{rCl}
        X_k^n & = & \tilde{f}_k^{(n)} \big( M_{k}^{(F)}, M_{k}^{(S)}, T_{k-1\to k }^{(1)}, \ldots, T_{k-1\to k}^{(\km)}, \nonumber \\
        & & \hspace{4cm} T_{k+1\to k }^{(1)}, \ldots, T_{k+1\to k}^{(\km)} \big)\IEEEeqnarraynumspace
        \end{IEEEeqnarray}
        subject to the power constraint in \eqref{eq:power}.
        
Each Rx~$(k)$ decodes the two messages $M_k^{(F)}, M_k^{(S)}$ only based on its channel outputs $Y_k^n$:
\begin{equation}
\big( \hat{M}_k^{(F)}, \hat{M}_{k}^{(S)}\big)=\tilde{g}^{(n)}_k\big( Y_k^n \big).
\end{equation}
Capacity region and  multiplexing gain region $\tilde{\mathcal{S}}^\star(\mu)$ are defined analogously as for receiver conferencing.

\begin{theorem}[Only Transmitter-Conferencing]\label{thm:txconf}
        Given $\mu \geq 0$, the per-user multiplexing gain  region $\tilde{\mathcal{S}}^\star(\mu)$ is the set of all nonnegative pairs $(\S^{(F)}, \S^{(S)})$ that satisfy
        \begin{align} \label{a1}
        2\S^{(F)} + \S^{(S)} & \le 1  \\
        \S^{(F)} + \S^{(S)} & \le \min\bigg\{  \frac{1}{2} + {\mu},\;\; \frac{ 2 \km +1}{2 \km+2} \bigg\}. \label{a2}
        \end{align}
\end{theorem}
\begin{IEEEproof}
Similar to the proof of Theorem~\ref{lemma3}. See Section~\ref{sec:txconf}. % Omitted due to space limitations.
        \end{IEEEproof}
        
%        To highlight a duality between transmitter and receiver-conferencing, we also consider the setup where only transmitters can conference but not the receivers.
%        \begin{theorem}
%        If only the transmitters can hold conferencing communications but not the receivers, the  multiplexing gain region is characterized by constraints \eqref{mg1} and \eqref{mg2} in Theorem~\ref{lemma3}.
%        \end{theorem}
%        \begin{IEEEproof}
%                Similar to the proof of Theorem~\ref{lemma3}. Omitted due to space limitations.
%        \end{IEEEproof}
        
        \begin{remark}
        Our results exhibit a duality between transmitter and receiver conferencing. They yield the same multiplexing gain region. %Moreover, for small values of $\mu$, this
        \end{remark}

 %%%%%%%%%%%%%%%%%%%
{\color{black}\section{Proof of Achievability of Theorem \ref{lemma1}}\label{sec:scheme}
%We present two coding schemes. Our first scheme achieves the rate-pairs in \eqref{eq:ach1}, and receivers conference only   parts of decoded ``fast" messages.  A single-round of conferencing suffices, and thus also ``slow" message are decoded with only two units of  delay. The second  scheme achieves the rate-pairs in \eqref{eq:ach2}, and receivers conference also parts of ``slow" messages. In general, here, $\km$ conferencing rounds are employed and ``slow" messages are decoded with maximum allowed delay $\km$. 
\subsection{Scheme 1: Conferencing only parts of ``fast"-messages}
Fix a small number $\epsilon>0$ and a joint distribution $P_{U_1 U_2 X}$ that satisfies the Markov chain $U_1 \to U_2 \to X$. Let $(U_1', U_2', X')$ be an independent copy of $(U_1,U_2,X)$ and define 
\begin{equation}
Y= X + \alpha X' +Z,
\end{equation}
where $Z$ is standard Gaussian independent of all other defined random variables.

Split each source message into two parts,  $M_k^{(F)} = \big(M_k^{(F_1)}, M_k^{(F_2)}\big)$,  of rates $\big(R_k^{(F_1)},R_k^{(F_2)}\big)$ that sum up to $R_k^{(F)} = R_k^{(F_1)} + R_k^{(F_2)}$ and so that 
\begin{equation}\label{eq:cons_conf}
R_k^{(F_1)} < \pi.
\end{equation}

\underline{\emph{Codebook construction:}} For each $k\in\{1,\ldots, K\}$,  generate  codebooks $\mathcal{C}_{1,k}$, $\{\mathcal{C}_{2,k}(i)\}$, and $\{\mathcal{C}_{x,k}(i,j)\}$ randomly. 
Codebook 
\begin{equation}
\mathcal{C}_{1,k}:= 
\Big \{ u_{1,k} ^ n(i) \colon \; \; i= 1, \ldots, \Big \lfloor 2^{n R_k^{(F_1)}} \Big \rfloor\Big\}
\end{equation} 
is generated by picking all entries i.i.d. according to $P_{U_1}$. 
For each $i\in\big\{1, \ldots, \big \lfloor 2^{n R_k^{(F_1)}}\big \rfloor \big\}$, codebook 
\begin{equation}
\mathcal{C}_{2,k}(i ) := \Big\{u_{2,k}^n(j|i) \colon\;  \; j= 1,\ldots,\Big \lfloor 2^{n R_k^{(F_2)}} \Big \rfloor \Big\}
\end{equation} is generated by picking the $t$-th entry of codeword $u_{2,k}^{n}(j|i)$  independently of all other entries and codewords according to the distribution $P_{U_2|U_1}(\cdot | u_{1,k,t}(i))$. Here, $u_{1,k,t}(i)$ denotes the $t$-th entry of codeword $u_{1,k}^n(i)$. 
For each pair $(i,j)$ in $\big\{1, \ldots, \big \lfloor 2^{n R_k^{(F_1)}}\big \rfloor \big\} \times \big\{1, \ldots, \big \lfloor 2^{n R_k^{(F_2)}}\big \rfloor \big\}$, codebook 
 \begin{equation}
 \mathcal{C}_{x,k}(i,j) := \Big\{x_k^n(\ell| i,j) \colon\;\; \ell = 1,\ldots,\Big \lfloor 2^{n R_k^{(S)}} \Big \rfloor \Big\}
 \end{equation}is generated by picking the $t$-th entry of codeword $x_k^{n}(\ell|i,j)$ independently of all other entries according to the distribution $P_{X|U_2}(\cdot | u_{2,k,t}(j|i))$. Here, $u_{2,k,t}(j|i)$ denotes the $t$-th entry of codeword $u_{2,k}^n(j|i)$. 

Reveal all codebooks to all terminals.

\underline{\emph{Encoding:}}  Tx~$(k)$ sends  codeword $x_{k}^n\big( M_{k}^{(S)} \big| M_{k}^{(F_1)}, M_{k}^{(F_2)}\big)$ over the channel.

\underline{\emph{Decoding:}} Each Rx~$(k)$ performs the following steps. Given that it observes $Y_k^n=y_k^n$, it first 
 looks for a unique pair $(\hat i,\hat j)$ such that  
\begin{equation}\label{eq:dec1}
(u_{1,k}^n(\hat i), u_{2,k}^n(\hat j|\hat i), y_k^n) \in T_{\epsilon}^{(n)}(P_{U_1U_2Y}).
\end{equation}
If none or more than one such pair $(\hat i,\hat j)$ exists, Rx~$(k)$ declares an error.  
Otherwise, it declares $\hat{M}_k^{(F)}= (\hat{i}, \hat{j})$, and it  sends 
\begin{equation}
Q_{k\to k+1}^{(1)}=\hat{i}.
\end{equation} to its right neighbour, Rx~$(k+1)$. 

With the message $Q_{k-1 \to k}^{(1)}$, Rx~$(k)$ obtains from its left-neighbour, it decodes also its intended ``slow" message. To this end, it looks for an index $\hat{\ell}$ such that 
\begin{align}\label{eq:dec2}
\Big(u_{1,k}^n(\hat i), u_{2,k}^n(\hat j|\hat i),x_k^n(\hat{\ell} |\hat i, \hat j),& \notag\\&\hspace{-2cm}u_{k-1}^n( Q_{k-1 \to k}^{(1)}), y_k^n\Big) \in T_{\epsilon}^{(n)}(P_{U_1U_2XU_1' Y}),
\end{align}
%where $P_{U_1U_2XU_1'Y}(u_1,u_2,x,u_1', y)= \sum_{u_2',x'} P_{U_1U_2X}(u_1,u_2,x) P_{U_1U_2X}(u_1', u_2',x') f_G(y-x-\alpha x')$ for $f_G$ denoting the standard Gaussian density.
If none or multiple such indices $\hat \ell$ exist, an error is declared.
Otherwise, Rx~$(k)$ declares $\hat{M}_{k}^{(S)} = \hat{\ell}$.

\underline{\emph{Analysis:}} 
Decoding in \eqref{eq:dec1} is successful with probability tending to 1 as $n \to \infty$, if 
\begin{align} \label{rf}
R_k^{(F_1)} + R_k^{(F_2)} & < I(U_{2}; Y) \\
R_k^{(F_2)} &< I(U_{2}; Y | U_{1}).\label{eq:rf2}
\end{align}

Decoding in \eqref{eq:dec2} is successful with probability tending to 1 as $n \to \infty$, if 
\begin{equation} \label{rs}
R_k^{(S)} < I(X; Y ,U_{1}'| U_{1}).
\end{equation}

The conferencing constraint is satisfied by \eqref{eq:cons_conf}.
Apply then Fourier-Motzkin elimination to  \eqref{eq:cons_conf}, \eqref{rf} and \eqref{eq:rf2}. Achievability of the pairs \eqref{eq:ach1} follows then by a rate-transfer argument that parts of the ``slow" messages   can also be sent as ``fast" messages.
%\end{IEEEproof}
%%%%%%%%%%%%%%%%%%%%%
\subsection{Scheme 2: Conferencing also parts of ``slow"-messages}
Fix $\epsilon>0$ and  a joint distribution $P_{U V_1 \ldots V_{\km - 1}X}$ satisfying the Markov chain $U \to V_1 \to V_2 \to \ldots V_{\km-1}\to X$ and the rate constraint
\begin{equation}
R_k^{(F)} < \pi.
\end{equation}
 Let $(U', V_1', V_2', \ldots, V_{\km-1}', X')$ be an independent copy of $(U, V_1, \ldots, V_{\km - 1},X)$ and define 
	\begin{equation}
	Y= X + \alpha X' +Z,
	\end{equation}
	where $Z$ is independent standard Gaussian.
	
Split $M_k^{(S)}$ into $\km $ parts,  $M_k^{(S)} = (M_k^{(S_1)},\ldots,$ $M_k^{(S_{\km})})$,  of rates $(R_k^{(S_1)},\ldots R_k^{(S_{\km})})$ that sum up to $R_k^{(S)} = \sum_{d=1}^{\km} R_k^{(S_d)}$ and satisfy
\begin{equation}\label{eq:cons_conf2}
R_k^{(F)} + \sum_{d=1}^{\km-1}  R_k^{(S_d)} < \pi.
\end{equation}

\par\underline{\emph{Codebook construction:}} For each $k\in\{1,\ldots, K\}$,  generate the following codebooks $\mathcal{C}_{1,k}$, $\{\mathcal{C}_{2,k}\}, \ldots,  \{\mathcal{C}_{\km,k}\}, \{\mathcal{C}_{x,k}\}$ randomly. 
Codebook 
\begin{equation}
\mathcal{C}_{1,k}:= 
\Big \{ u_{k} ^ n(j) \colon j= 1, \ldots, \Big \lfloor 2^{n R_k^{(F)}} \Big \rfloor\Big\}
\end{equation} 
is generated by picking all entries i.i.d. according to $P_{U}$. 
For each $j\in\big\{1, \ldots, \big \lfloor 2^{n R_k^{(F)}}\big \rfloor \big\}$, codebook 
\begin{equation}
\mathcal{C}_{2,k}(j) := \Big\{v_{1,k}^n(i_1|j) \colon i_1= 1,\ldots,\Big \lfloor 2^{n R_k^{(S_1)}} \Big \rfloor \Big\}
\end{equation} is generated by picking the $t$-th entry of codeword $v_{1,k}^n(i_1|j)$   independently of all other entries and codewords according to the distribution $P_{V_{1,k}|U}(\cdot | u_{k,t}(i))$. Here, $u_{k,t}(i)$ denotes the $t$-th entry of codeword $u_{k}^n(i)$. 
For each $d \in \{2,\ldots,{\km-1}\}$, and  each tuple $(j,i_1,\ldots, i_{d-1})$ in $\big\{1, \ldots, \big \lfloor 2^{n R_k^{(F)}}\big \rfloor \big\} \times\prod_{\ell=1}^{d-1} \big\{1, \ldots, \big \lfloor 2^{n R_k^{(S_\ell)}}\big \rfloor \big\}$,
  codebook 
 \begin{align}
 \mathcal{C}&_{d + 1,k}(j,i_1,\ldots,i_{d-1})  \notag\\&:=\Big\{v_{d,k}^n(i_d| j,i_1,\ldots,i_{d-1}) \colon i_d = 1,\ldots,\Big \lfloor 2^{n R_k^{(S_{d})}} \Big \rfloor \Big\}
 \end{align}is generated by picking the $t$-th entry of codeword $v_{d,k}^n(i_d| j,i_1,\ldots,i_{d-1})$   independently of all other entries according to the distribution $P_{V_{d,k}|V_{d-1,k}}(\cdot | v_{d-1,k,t}(i_{d-1}|j,i_1,\ldots,i_{d-2}))$. Here, $v_{d-1,k,t}(i_{d-1}|j,i_1,\ldots,i_{d-2})$ denotes the $t$-th entry of codeword $v_{d-1,k}^n(i_{d-1}|j,i_1,\ldots,i_{d-2})$. 
 
 Finally, for each 
tuple $(j,i_1,\ldots, i_{\km-1})$ in $\big\{1, \ldots, \big \lfloor 2^{n R_k^{(F)}}\big \rfloor \big\} \times\prod_{\ell=1}^{\km-1} \big\{1, \ldots, \big \lfloor 2^{n R_k^{(S_\ell)}}\big \rfloor \big\}$,
codebook 
\begin{align}
\mathcal{C}&_{x,k}(j,i_1,\ldots,i_{\km-1})  \notag\\&:=\Big\{x_{k}^n(\ell| j,i_1,\ldots,i_{\km-1}) \colon \;\; \ell  = 1,\ldots,\Big \lfloor 2^{n R_k^{(S_{\km})}} \Big \rfloor \Big\}
\end{align}is generated by picking the $t$-th entry of codeword $x_{k}^n(\ell| j,i_1,\ldots,i_{\km-1})$   independently of all other entries according to the distribution $P_{X_{k}|V_{\km-1,k}}(\cdot | v_{\km-1,k,t}(i_{\km-1}|j,i_1,\ldots,i_{\km-2}))$. Here, $v_{\km-1,k,t}(i_{\km-1}|j,i_1,\ldots,i_{\km-2})$ denotes the $t$-th entry of codeword $v_{\km-1,k}^n(i_{\km-1}|j,i_1,\ldots,i_{\km-2})$.
Reveal all codebooks to all terminals.

\underline{\emph{Encoding:}}  Tx~$(k)$ sends  codeword $$x_{k}^n\big( M_{k}^{(S_{\km})} | M_{k}^{(F)},\ldots, M_{k}^{(S_{\km-2})},M_{k}^{(S_{\km-1})} \big)$$
\underline{\emph{Decoding:}}  Each Rx~$(k)$ performs the following steps. Given that it observes $Y_k^n=y_k^n$, it first 
 looks for a unique index $\hat j$ such that  
\begin{equation}\label{eq:dee0}
(u_{k}^n(\hat j), y_k^n) \in T_{\epsilon}^{(n)}(P_{UY}).
\end{equation}
If none or more than one such index $\hat j$ exist, Rx~$(k)$ declares an error.  
Otherwise, it declares $\hat{M}_k^{(F)}= \hat{j}$, and  sends 
\begin{equation}
Q_{k\to k+1}^{(1)}=\hat{j}
\end{equation} to its right neighbour, Rx~$k+1$. 

With the message $Q_{k-1 \to k}^{(1)}$, Rx~$(k)$ obtains from its left-neighbour,  it looks for an index $\hat{i_1}$ such that 
\begin{align}
\Big(u_{k}^n(\hat j), v_{1,k}^n(\hat i_1|\hat j),u_{k-1}^n( Q_{k-1 \to k}^{(1)}), y_k^n\Big) \in T_{\epsilon}^{(n)}(P_{UV_1U'Y}). \label{eq:dee1}
\end{align}
If none or multiple such index  $\hat i_1$ exist, Rx~$(k)$ declares an error.  
Otherwise, it declares $\hat{M}_k^{(S_1)}= \hat{i_1}$ and sends 
\begin{equation}
Q_{k\to k+1}^{(2)}=\hat{i_1}
\end{equation} to its right neighbour, Rx~$(k+1)$. 

In conferencing round $d$ with $d \in\{3,\ldots,\km-1\}$, Rx~$(k)$ obtains $Q_{k-1 \to k}^{(d)}$ from its left-neighbour, and looks for an index $\hat i_{d}$  %with $d \in \{2,\ldots,\km-1\} $ such that 
\begin{align}
\Big(u_{k}^n(\hat j)&, v_{1,k}^n(\hat i_1|\hat j),\ldots,v_{d,k}^n(\hat i_{d}|\hat j,\hat  i_1,\ldots,\hat i_{d-1}), u_{k-1}^n( Q_{k-1 \to k}^{(1)}), \notag\\ &\hspace{1.4cm} v_{1,k-1}^n( Q_{k-1 \to k}^{(2)}),\ldots ,v_{d-1,k-1}^n( Q_{k-1 \to k}^{(d)}), y_k^n\Big) \notag\\&\hspace{1cm}\in T_{\epsilon}^{(n)}(P_{UV_1\ldots V_{d}U'V_1'\ldots V'_{d}Y}). \label{eq:dee2}
\end{align}
%where 
%\begin{align}
%&P_{UV_1\ldots V_{d}U'V_1'\ldots V'_{d-1}Y}(u,v_1,\ldots, v_d,u',v'_1,\ldots,v'_{d-1}, y)\notag\\&= \sum_{v'_d} \Big(P_{UV_1\ldots V_{d}}(u,v_1,\ldots, v_d) P_{UV_1\ldots V_{d}}(u',v'_1,\ldots,v'_{d}) \notag\\&\hspace{3cm}f_G(y-v_{d}-\alpha v'_{d})\Big),
%\end{align} for $f_G$ denoting the standard Gaussian density.
If none or more than one such index $\hat i_d$  exist, an error is declared.
Otherwise, Rx~$(k)$ declares $\hat{M}_{k}^{(S_\ell)} =\hat i_d$ and sends
\begin{equation}
Q_{k\to k+1}^{(d)}=\hat{i_d}
\end{equation} to its right neighbour, Rx~$(k+1)$.
	
In conferencing round $\km$,  Rx~$(k)$ obtains $Q_{k-1\to k}^{(\km-1)}$ from its left-neighbour and looks for an index $\hat \ell$ 
\begin{align}
\Big(u_{k}^n(\hat j)&, v_{1,k}^n(\hat i_1|\hat j),\ldots,v_{\km-1,k}^n(\hat i_{\km-1}|\hat j,\hat  i_1,\ldots,\hat i_{\km-2}),  \notag\\ &\hspace{-0.5cm} x_{k}^n(\hat i_{\km}|\hat j,\hat  i_1,\ldots,\hat i_{\km-1}),u_{k-1}^n( Q_{k-1 \to k}^{(1)}),\notag\\&v_{1,k-1}^n( Q_{k-1 \to k}^{(2)}),\ldots ,v_{\km-1,k-1}^n( Q_{k-1 \to k}^{(\km-1)}), y_k^n\Big) \notag\\&\hspace{1cm}\in T_{\epsilon}^{(n)}(P_{UV_1\ldots V_{\km-1}XU'V_1'\ldots V'_{\km-1}X'Y}). \label{eq:dee3}
\end{align}
If none or multiple such index $\hat \ell$ exist, an error is declared.
Otherwise, Rx~$(k)$ declares $\hat{M}_{k}^{(S_{\km})} = \hat{\ell}$.

\underline{\emph{Analysis:}} 
Decoding in \eqref{eq:dee0} is successful with probability tending to 1 as $n \to \infty$, if 
\begin{equation} \label{rf21}
R_k^{(F)}  < I(U; Y) 
\end{equation}
Decoding in \eqref{eq:dee1} is successful with probability tending to 1 as $n \to \infty$, if 
\begin{equation} \label{rs21}
R_k^{(S_1)} < I(V_{1}; Y, U'| U).
\end{equation}
Decoding in \eqref{eq:dee2} is successful with probability tending to 1 as $n \to \infty$, if 
\begin{equation} \label{rs31}
R_k^{(S_d)} < I(V_{d}; Y, V_{d-1}'| V_{d-1}), \qquad d \in \{2,\ldots, \km-1\}.
\end{equation}
The last decoding step in \eqref{eq:dee3} is successful with probability tending to 1 as $n \to \infty$, if 
\begin{equation} \label{rs32}
R_k^{(S_{\km})} < I(X; Y, X'| V_{\km-1}).
\end{equation} 
The conferencing constraint is satisfied by \eqref{eq:cons_conf2}. 
Apply Fourier-Motzkin elimination to \eqref{eq:cons_conf2}, \eqref{rf21}, \eqref{rs21},\eqref{rs31} and \eqref{rs32}.  Achievability of the pairs \eqref{eq:ach2} follows then by a rate-transfer argument.}
%
%
%
%
%

 %%%%%%%%%%%%%%%%%%%%%%%%%%%%%%
 \section{Proof of Theorem~\ref{thm:converse}}\label{sec:converse}

%\subsection{ For $\mu < \kappa_{\text{max}}/ (\kappa_{\text{max}} + 2)$ }
For convenience of notation, define for any $k\in \{1,\ldots, K\}$:
\begin{equation}
M_k := (\MkF,\MkS).
\end{equation}
We first prove Inequality~\eqref{eq:conv1}. By Fano's Inequality and the independence of the messages, we have for any $k \in \{1,\ldots, K-1\}$:
\begin{IEEEeqnarray}{rCl}
\lefteqn{R_k^{(F)} + R_k^{(S)} + R_{k+1}^{(F)} } \nonumber \\
 & = &\hspace{-0.15cm} \frac{1}{n} \Big[ H(M_k^{(F)}) + H(M_k^{(S)}) + H(M_{k+1}^{(F)})\Big]\notag\\
  & = & \hspace{-0.15cm}\frac{1}{n} \Big[ H(M_k^{(F)} |M_{k-1}) \nonumber \\
  && \hspace{.7cm}   + H(M_k^{(S)} | M_1,\ldots,M_{k-1}, M_k^{(F)}, M_{k+1}, \ldots, M_{K} )  \nonumber \\
&& \qquad \hspace{3.8cm}  + H(M_{k+1}^{(F)}|M_{k - 1}, M_{k+1}^{(S)})\Big]\notag\\
& \le&\hspace{-0.1cm} \frac{1}{n} \Big[ I(M_k^{(F)}; Y_k^{n} | M_{k - 1})  \notag \\ &  & \qquad + I(M_k^{(S)}; Y_1^{n}, \ldots, Y_K^{n}|M_1,\ldots,M_{k-1}, M_k^{(F)}\notag\\
& &\hspace{4.7cm} ,M_{k+1}, \ldots, M_{K}) \notag \\  & &\hspace{3cm} + I(M_{k+1}^{(F)}; Y_{k+1}^{n} |  M_{k - 1}, M_{k+1}^{(S)}) \Big]   \notag \\  & &\hspace{7.1cm}+ \frac {\epsilon_n}{n}\notag \\ 
& \stackrel{(a)}{=}& \frac{1}{n} \Big[ I(M_k^{(F)}; Y_k^{n} |  M_{k - 1})  \notag \\ 
&  & \qquad + I(M_k^{(S)}; Y_k^{n}, Y_{k+1}^n| M_{k-1}, M_k^{(F)}, M_{k+1}) \nonumber  \\ 
 & &\hspace{2.4cm} + I(M_{k+1}^{(F)}; Y_{k+1}^{n} | M_{k - 1}, M_{k+1}^{(S)}) \Big]  + \frac {\epsilon_n}{n}\notag \\ 
& \stackrel{(b)}{=} &\frac{1}{n} \Big[I(M_k^{(F)}, M_k^{(S)}; Y_k^{n} |  M_{k - 1}) \notag\\ 
&&\qquad + I(M_k^{(S)}; Y_{k+1}^{n} | Y_k^{n}, M_k^{(F)}, M_{k - 1}, M_{k +1})\notag\\
 &&\qquad \hspace{1.7cm}+ I(M_{k+1}^{(F)}; Y_{k+1}^{n} |  M_{k - 1}, M_{k+1}^{(S)}) \Big] + \frac {\epsilon_n}{n} \notag \\ 
 &\leq & \hspace{-0.15cm}\frac{1}{n} \Big[ h(X_k^n + Z_k^n) - h(Z_k^n) + h(\alpha X_k^n + Z_{k+1}^n|  X_k^n +  Z_{k}^n) \notag\\  
 &&\quad \quad-h(Z_{k+1}^n)+ h(Y_{k + 1}^ n| M_{k+1}^{(S)}) - h(\alpha X_k^n + Z_{k+1}^n) \Big] \nonumber \\
 && \hspace{7.1cm}+ \frac {\epsilon_n}{n}\notag\\
 & \stackrel{(c)}{\leq }  & %& \frac{1}{n} \big[ h(X_k^n + Z_k^n) - h(\alpha X_k^n + Z_{k}^n) \notag \\
 %& & \qquad + I(Y_{k+1}^n; X_{k+1}^n, X_k^n) +  h(Z_{k+1}^n - \alpha Z_k^n) - h(Z_{k+1}^n)\big] \notag\\ 
 \frac{1}{2} \log (1+(1+ |\alpha|^2) P) + \frac{1}{2} \log (1+ \alpha^2) \nonumber\\
 & & +\max \{ -  \log |\alpha| , 0\} + \frac {\epsilon_n}{n}. \label{eq:b1}
\end{IEEEeqnarray}
Here, $(a)$ follows because given source messages $M_{k-1}$ and $M_{k+1}$, the triple  $(M_k, Y_k^n, Y_{k+1}^n)$ is independent of  the rest of the outputs $Y_{1}^n, \ldots, Y_{k-1}^n, Y_{k+2}^n, \ldots, Y_{K}^n$ and  source messages $M_1, \ldots, M_{k-2}, M_{k+2}, \ldots, M_K$; $(b)$ follows by the chain rule of mutual information and because $M_{k+1}$ is independent of the tuple $(M_{k-1}, M_k, Y_k^n)$; $(c)$  is obtained by  rearranging terms, and  the following  bounds \eqref{ee1}--\eqref{ee4}. In fact, because conditioning can only reduce entropy, and by the entropy-maximizing property of the Gaussian distribution,
\begin{IEEEeqnarray}{rCl}\label{ee1}
	h(Y_{k+1}^n |M_{k+1}^{(S)}) & \leq & h(Y_{k+1}^n )  \nonumber \\
	&\leq & \frac{1}{2} \log ((2\pi e)(1+(1+ |\alpha|^2) P)). \IEEEeqnarraynumspace
	\end{IEEEeqnarray}
Moreover, 
	\begin{IEEEeqnarray}{rCl}
		h(\alpha X_{k}^n + Z_{k+1}^n | X_k^n +Z_k^n)& =& h( Z_{k+1}^n-\alpha Z_{k}^n | X_k^n +Z_k^n) \nonumber \\
		 & \leq & h( Z_{k+1}^n-\alpha Z_{k}^n )\nonumber \\
		& = &  \frac{1}{2} \log ((2\pi e)(1+ \alpha^2)).
	\end{IEEEeqnarray}
	For the next bound, define $T_k^n$ i.i.d. zero-mean Gaussian independent of all other random variables and with a variance that depends on $\alpha$. If $\alpha<1$, the variance is $\frac{1}{\alpha^2}-1$. In this case,  $\frac{1}{\alpha} Z_{k+1}^n$ has the same joint distribution  with all other random variables as $Z_k^n+ T_k^n$ and
	\begin{IEEEeqnarray}{rCl} \lefteqn{ h(X_k^n + Z_k^n) -  h(\alpha X_k^n + Z_{k+1}^n) } \qquad  \nonumber \\
		& = & h(X_k^n + Z_k^n) -  h( X_k^n + \frac{1}{\alpha} Z_{k+1}^n)- \log |\alpha| \nonumber \\
				& = & h(X_k^n + Z_k^n) -  h( X_k^n + Z_{k}^n +T_k^n)- \log |\alpha| \nonumber \\
		& \leq & - \log |\alpha|.\label{ee2}
	\end{IEEEeqnarray}
	If $\alpha \geq 1$, then each symbol of $T_k^n$ has  variance $1-\frac{1}{\alpha^2}$. In this case,  $Z_k^n$ has the same joint distribution with all other random variables as $\frac{1}{\alpha} Z_{k+1}^n+ T^n_k$. Thus,  similarly to before:
		\begin{IEEEeqnarray}{rCl} \lefteqn{ h(X_k^n + Z_k^n) -  h(\alpha X_k^n + Z_{k+1}^n) } \;  \nonumber \\
			& = & h(X_k^n + \frac{1}{\alpha} Z_{k+1}^n+ T^n_k) -  h( \alpha X_k^n + Z_{k+1}^n )\nonumber\\
						& = & h(X_k^n + \frac{1}{\alpha} Z_{k+1}^n+ T^n_k) -  h( X_k^n +  \frac{1}{\alpha }Z_{k+1}^n |T^n_k)- \log |\alpha| \nonumber\\
			& \leq &  I(X_k^n  + \frac{1}{\alpha} Z_{k+1}^n+ T^n_k ; T_k^n)- \log |\alpha| \nonumber \\
			 &  \leq & I( \frac{1}{\alpha} Z_{k+1}^n+ T^n_k ; T_k^n)- \log |\alpha| \nonumber \\
			 &= &  \frac{1}{2} \log \Big(  \frac{1}{1/\alpha^2}\Big) - \log |\alpha| =0.\label{ee4}
		\end{IEEEeqnarray}

Following similar steps, one can also prove that 
\begin{IEEEeqnarray}{rCl}\label{eq:b2}
R_K^{(F)} + R_K^{(S)} & \leq& \frac{1}{n}  I(M_k^{(F)}, M_k^{(S)}; Y_k^{n} |  M_{k - 1})  +\frac{\epsilon_n}{n}  \nonumber \\ 
& \leq &  \frac{1}{2} \log (1+P) + \frac{\epsilon_n}{n}.
\end{IEEEeqnarray}

We  sum up  the  bound in \eqref{eq:b1}   for all  values of $k\in\{1,\ldots, K-1\}$,  and combine it with \eqref{eq:b2}. Taking  $n \to \infty$,  it follows  that whenever the probability of error $P_e^{(n)}$ vanishes as $n\to \infty$ (and thus $\frac{\epsilon_n}{n}\to 0$ as $n\to \infty$):
\begin{IEEEeqnarray}{rCl} \label{bn1}
\lefteqn{\sum_{k = 1}^K \left(2 R_k^{(F)} + R_k^{(S)}\right) }\nonumber \\
& = & R_1^{(F)} + \sum_{k=1}^{K-1} \Big(R_k^{(F)}+ R_k^{(S)}+ R_{k+1}^{(F)}\Big) + R_K^{(F)} + R_K^{(S)}    \notag\\
& \leq &  (K-1)\frac{1}{2} \log (1+(1+ \alpha^2) P)  + \log( 1+P)  \nonumber\\
& & + \frac{K-1}{2} \log (1+ \alpha^2) +(K-1)  \max \{ -  \log |\alpha| , 0\},\IEEEeqnarraynumspace
\end{IEEEeqnarray}

{\color{black}We now  prove bound \eqref{eq:conv2}. We assume $K$ is even. For $K$ odd the bound can be proved in a similar way. Recall that $X_0^n=0$, and define 
\begin{IEEEeqnarray}{rCl}
	\mathbf{M}_{\textnormal{odd}} &:=& \{M_k \colon \;\;  k \textnormal{ odd}\}\nonumber\\
		\mathbf{M}_{\textnormal{even}} &:=& \{M_k \colon \;\;  k \textnormal{ even}\}\nonumber\\
					\mathbf{X}^n_{\textnormal{odd}} &:=& \{X_k^n \colon k \textnormal{ odd}\}\nonumber\\
					\mathbf{X}^n_{\textnormal{even}} &:=& \{X_k^n \colon \;\;  k \textnormal{ odd}\}\nonumber\\
			\mathbf{Y}^n_{\textnormal{odd}} &:=& \{Y_k^n \colon k \textnormal{ odd}\}\nonumber\\
			\mathbf{Y}^n_{\textnormal{even}} &:=& \{Y_k^n \colon k \textnormal{ even}\}\nonumber\\
				\mathbf{Z}^n_{\textnormal{odd}} &:=& \{Z_k^n \colon k \textnormal{ odd}\}\nonumber\\
				\mathbf{Z}^n_{\textnormal{even}} &:=& \{Z_k^n \colon k \textnormal{ even}\}\nonumber
	\end{IEEEeqnarray}
	and
\begin{IEEEeqnarray}{rCl}
	\mathbf{Q}_{\textnormal{odd}} & :=& \Big\{Q_{k\to \tilde{k}}^{(1)},\ldots, Q_{k\to \tilde{k}}^{(\km)} \colon  \ k \textnormal{ odd },\ \tilde{k}\in\{k-1, k+1\}\Big\}\nonumber \\
		\mathbf{Q}_{\textnormal{even}} & :=& \Big\{Q_{k\to \tilde{k}}^{(1)},\ldots, Q_{k\to \tilde{k}}^{(\km)}\colon \ k \textnormal{ even}, \ \tilde{k}\in\{k-1, k+1\}\Big\}. \nonumber 
\end{IEEEeqnarray}
By Fano's inequality, there must exist a sequence $\{\epsilon_n\}_{n=1}^\infty$ so that $\frac{\epsilon_n}{n} \to 0$ as $n\to \infty$ and 
\begin{IEEEeqnarray}{rCl}
\lefteqn{	\sum_{k=1}^K \left( R_k^{(F)} +R_k^{(S)} \right)} \nonumber \\
& =&\hspace{-0.2cm} \frac{1}{n} \big[H(\mathbf{M}_{\textnormal{odd}}) + H(\mathbf{M}_{\textnormal{even}}) \big] \nonumber \\
 & \leq & \hspace{-0.2cm}\frac{1}{n} \big[I(\mathbf{M}_{\textnormal{odd}}; \mathbf{Y}_{\textnormal{odd}}, \mathbf{Q}_{\textnormal{odd}} ) + I (\mathbf{M}_{\textnormal{even}}; \mathbf{Y}_{\textnormal{even}} ,\mathbf{Q}_{\textnormal{even}} | \mathbf{M}_{\textnormal{odd}}  ) \big] \nonumber \\
 & &+ \frac{\epsilon_n}{n} \nonumber \\
 & = &\hspace{-0.2cm}  \frac{1}{n} \big[I(\mathbf{M}_{\textnormal{odd}}; \mathbf{Y}_{\textnormal{odd}} ) + I (\mathbf{M}_{\textnormal{even}}; \mathbf{Y}_{\textnormal{even}} | \mathbf{M}_{\textnormal{odd}}  ) \nonumber \\
 & & \hspace{2mm} + I(\mathbf{M}_{\textnormal{odd}};  \mathbf{Q}_{\textnormal{odd}} |\mathbf{Y}_{\textnormal{odd}} ) + I (\mathbf{M}_{\textnormal{even}}; \mathbf{Q}_{\textnormal{even}} |  \mathbf{M}_{\textnormal{odd}}, \mathbf{Y}_{\textnormal{even}}   ) \big] \nonumber\\ 
 & &+ \frac{\epsilon_n}{n} \nonumber \\
 & \leq &\hspace{-0.25cm}  \frac{1}{n} \big[h(\mathbf{Y}_{\textnormal{odd}} )  - h(\mathbf{Y}_{\textnormal{odd}} | \mathbf{M}_{\textnormal{odd}})  + h( \mathbf{Y}_{\textnormal{even}} | \mathbf{M}_{\textnormal{odd}}  )  - h(\mathbf{Z}_{\textnormal{even}} )\big] \nonumber \\
 & & \hspace{5mm} + H( \mathbf{Q}_{\textnormal{odd}}) + I (\mathbf{M}_{\textnormal{even}}; \mathbf{Q}_{\textnormal{even}} |  \mathbf{M}_{\textnormal{odd}}, \mathbf{Y}_{\textnormal{even}}   ) \big] + \frac{\epsilon_n}{n}\nonumber\\ 
 & \leq  & \hspace{-0.2cm} \frac{1}{n} \big[h(\mathbf{Y}_{\textnormal{odd}} )  - h(\mathbf{Y}_{\textnormal{odd}} | \mathbf{M}_{\textnormal{odd}})  \nonumber \\
 & & \hspace{5mm}  +h(\mathbf{Y}_{\textnormal{even}} | \mathbf{M}_{\textnormal{odd}})   - h(\mathbf{Z}_{\textnormal{even}} )+ H( \mathbf{Q}_{\textnormal{odd}}) \nonumber \\
 & &\hspace{10mm} + I (\mathbf{M}_{\textnormal{even}}; \mathbf{Q}_{\textnormal{even}} |  \mathbf{M}_{\textnormal{odd}}, \mathbf{Y}_{\textnormal{even}} , \mathbf{Z}_{\textnormal{even}} - \alpha^{-1}  \mathbf{Z}_{\textnormal{odd}}  )  \nonumber\\
 & & \hspace{18mm} +  I (\mathbf{M}_{\textnormal{even}};   \mathbf{Z}_{\textnormal{even}} - \alpha^{-1}  \mathbf{Z}_{\textnormal{odd}}  | \mathbf{M}_{\textnormal{odd}}, \mathbf{Y}_{\textnormal{even}}  ) \big] \nonumber \\ &+ \frac{\epsilon_n}{n}\nonumber\\ 
 & \stackrel{(a)}{\leq}  &\Big(\frac{K}{2} +1\Big) \cdot\frac{1}{2} \log (1 + (1+\alpha^2)P)  \nonumber \\
  && +     \frac{K-2}{2} \cdot  \max \{-\log|\alpha|, 0 \}\nonumber \\
 & & \hspace{5mm} + \pi K + \frac{K}{2}  \frac{1}{2} \log ( 1 +\alpha^2) + \frac{\epsilon_n}{n}, 
\end{IEEEeqnarray} 
where we 
where $(a)$ holds because:
\begin{itemize}
	\item By the entropy maximizing property of the Gaussian distribution:
	\begin{equation}
h(\mathbf{Y}_{\textnormal{odd}}) - h(\mathbf{Z}_{\textnormal{even}}) \leq n % \left %\lceil\frac{K}{2}\right \rceil
\frac{K}{2} \cdot\frac{1}{2} \log (1 + (1+\alpha^2)P);
	\end{equation}
	\item By \eqref{ee2} and \eqref{ee4}: 
	\begin{IEEEeqnarray}{rCl}
	\lefteqn{	h(\mathbf{Y}_{\textnormal{odd}} | \mathbf{M}_{\textnormal{odd}}) - h(\mathbf{Y}_{\textnormal{even}} | \mathbf{M}_{\textnormal{odd}}) }\nonumber  \qquad \\
	& = &h(Y_K^n| M_{k-1})- h(Y_1^n|M_1) \nonumber \\
	& & + \sum_{i=1}^{K/2-1}  \big[h( X_{2i}^n+Z_{2i}^n)- h(\alpha X_{2i}^n + Z_{2i+1}^n) \big] \nonumber \\
	&  \leq &  n \frac{1}{2} \log (1 +(1+\alpha^2) P) \nonumber \\
	& & + n \left( \frac{K}{2}-1\right) \max\{-\log|\alpha|,0\} ; \end{IEEEeqnarray}
	\item By the rate-limitation of the conferencing links:
	\begin{equation}
H(\mathbf{Q}_{\textnormal{odd}}) \leq n  \pi  K;	
\end{equation}
	\item From  the tuple $(\mathbf{M}_{\textnormal{odd}}, \mathbf{Y}_{\textnormal{even}} , \mathbf{Z}_{\textnormal{even}} - \alpha^{-1}  \mathbf{Z}_{\textnormal{odd}})$ it is possible to compute also $\mathbf{Y}_{\textnormal{odd}}$ and thus $\mathbf{Q}_{\textnormal{even}} $:
		\begin{equation}I(\mathbf{M}_{\textnormal{even}}; \mathbf{Q}_{\textnormal{even}} |  \mathbf{M}_{\textnormal{odd}}, \mathbf{Y}_{\textnormal{even}} , \mathbf{Z}_{\textnormal{even}} - \alpha^{-1} \mathbf{Z}_{\textnormal{odd}}  )=0;
			\end{equation}
	\item By the fact that conditioning reduces entropy:
	 \begin{IEEEeqnarray}{rCl}\lefteqn{
	 I (\mathbf{M}_{\textnormal{even}};   \mathbf{Z}_{\textnormal{even}} - \alpha^{-1}  \mathbf{Z}_{\textnormal{odd}}  | \mathbf{M}_{\textnormal{odd}}, \mathbf{Y}_{\textnormal{even}}  )} \nonumber \\
	 & \leq &  h( \mathbf{Z}_{\textnormal{even}} - \alpha^{-1} \mathbf{Z}_{\textnormal{odd}})- 
h ( \mathbf{Z}_{\textnormal{even}} - \alpha^{-1} \mathbf{Z}_{\textnormal{odd}} |\mathbf{Z}_{\textnormal{even}})  \nonumber \\
&=& n \frac{K}{2} \cdot % \left \lceil \frac{K}{2} \right \rceil 
\frac{1}{2} \log ({ 1 +\alpha^2})
\end{IEEEeqnarray} 
\end{itemize} 
Taking $n \to \infty$ establishes the proof.
}
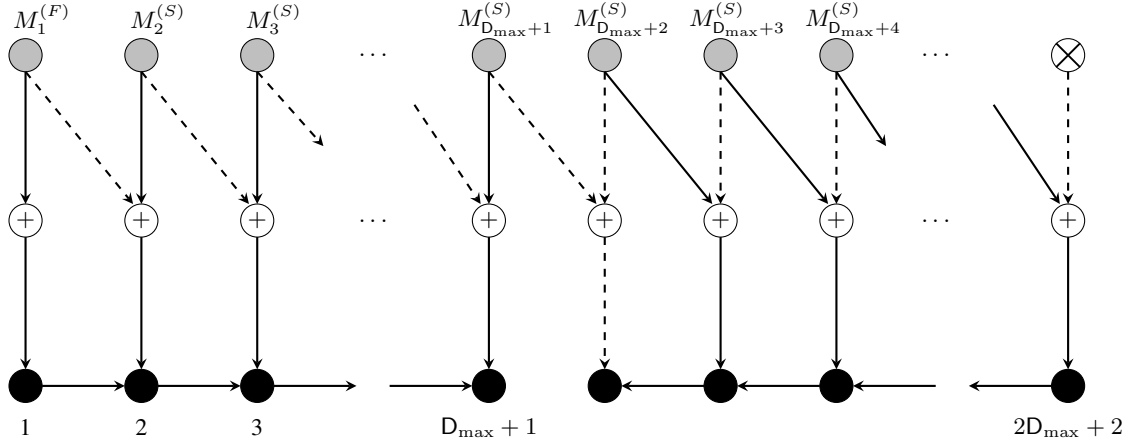
\begin{figure*}[t]
%\small
  \centering
  \small
    %\hspace*{32pt}
 \begin{tikzpicture}[scale=2.2, >=stealth]
\centering
\tikzstyle{every node}=[draw,shape=circle, node distance=0.5cm];
\draw [fill=gray!50](-2.5, 2) circle (0.1);
\draw [fill=gray!50](-1.8, 2) circle (0.1);
\draw [fill=gray!50](-1.1, 2) circle (0.1);
\draw [fill=gray!50](0.3, 2) circle (0.1);
\draw [fill=gray!50](1, 2) circle (0.1);
\draw [fill=gray!50](1.7, 2) circle (0.1);
\draw [fill=gray!50](2.4, 2) circle (0.1);
\draw [fill=white](3.8, 2) circle (0.1);
\node[draw =none, rotate =45] (s2) at (3.8,2) {\huge$+$};
%%%%%%%%%%%%%%
\draw [fill=black](-2.5, 0) circle (0.1);
\draw [fill=black](-1.8, 0) circle (0.1);
\draw [fill=black](-1.1, 0) circle (0.1);
\draw [fill=black](0.3, 0) circle (0.1);
\draw [fill=black](1,0) circle (0.1);
%\node[draw =none, rotate =45] (s2) at (1,0) {\huge$+$};
\draw [fill=black](1.7,0) circle (0.1);
\draw [fill=black](2.4, 0) circle (0.1);
\draw [fill=black](3.8, 0) circle (0.1);
%%%%%%%%%%%%%%%%%%%
\node[draw =none] (s2) at (-2.5,1 ) {$+$};
\draw (-2.5, 1) circle (0.1);
\node[draw =none] (s2) at (-1.8,1 ) {$+$};
\draw (-1.8,1) circle (0.1);
\node[draw =none] (s3) at (-1.1,1 ) {$+$};
\draw (-1.1, 1) circle (0.1);
\node[draw =none] (s4) at (0.3,1 ) {$+$};
\draw (0.3, 1) circle (0.1);
\node[draw =none] (s5) at (1,1 ) {$+$};
\draw (1,1) circle (0.1);
\node[draw =none] (s6) at (1.7,1 ) {$+$};
\draw (1.7,1) circle (0.1);
\node[draw =none] (s7) at (2.4,1 ) {$+$};
\draw (2.4,1) circle (0.1);
\node[draw =none] (s8) at (3.8,1 ) {$+$};
\draw (3.8,1) circle (0.1);
%%%%%%%%%%%%%%%%%%%%%%%%%%%%%
\draw   [thick,->] (-2.4,0)-- (-1.9,0);
\draw   [thick,->] (-1.7,0)-- (-1.2,0);
\draw   [thick,->] (-1,0)-- (-0.5,0);
\draw   [thick,->] (-0.3,0)-- (0.2,0);
%\draw   [thick,->] (0.4,0)-- (0.9,0);
\draw   [thick,->] (1.6,0)-- (1.1,0);
\draw   [thick,->] (2.3,0)-- (1.8,0);
\draw   [thick,->] (3,0)-- (2.5,0);
\draw   [thick,->] (3.7,0)-- (3.2,0);
%%%%%%%%%%%%%%%%%%%%%%%%%%%%%
\node[draw =none]  at (-2.4,2.22) {\small$M_1^{(F)}$};
\node[draw =none]  at (-1.7,2.22) {\small$M_2^{(S)}$};
\node[draw =none]  at (-1,2.22) {\small$M_3^{(S)}$};
\node[draw =none]  at (0.4,2.22) {\small$M_{\km+1}^{(S)}$};
\node[draw =none]  at (1.1,2.22) {\small$M_{\km+2}^{(S)}$};
\node[draw =none]  at (1.8,2.22) {\small$M_{\km+3}^{(S)}$};
\node[draw =none]  at (2.5,2.22) {\small$M_{\km+4}^{(S)}$};
%%%%%%%%%%%%%%%%%%%%%%%%%%%%
\node[draw =none]  at (-2.5,-0.25) {\small{1}};
\node[draw =none]  at (-1.8,-0.25) {\small2};
\node[draw =none]  at (-1.1,-0.25) {\small3};
\node[draw =none]  at (0.3,-0.25) {\small{$\km+1$}};
\node[draw =none]  at (3.8,-0.25) {\small{$2\km+2$}};
%%%%%%%%%%%%%%%%%%%%%%%%%%%%%%
\draw   [thick,->] (-2.5,1.9)-- (-2.5,1.1);
\draw   [thick,->] (-1.8,1.9)-- (-1.8,1.1);
\draw   [thick,->] (-1.1,1.9)-- (-1.1,1.1);
\draw   [thick,->] (0.3,1.9)-- (0.3,1.1);
\draw   [thick,->,dashed] (1,1.9)-- (1,1.1);
\draw   [thick,->,dashed] (1.7,1.9)-- (1.7,1.1);
\draw   [thick,->,dashed] (2.4,1.9)-- (2.4,1.1);
\draw   [thick,->,dashed] (3.8,1.9)-- (3.8,1.1);
%%%%%%%%%%%%%%%%%%%%%%%
\draw   [thick,->] (-2.5,0.9)-- (-2.5,0.1);
\draw   [thick,->] (-1.8,0.9)-- (-1.8,0.1);
\draw   [thick,->] (-1.1,0.9)-- (-1.1,0.1);
\draw   [thick,->] (0.3,0.9)-- (0.3,0.1);
\draw   [thick,->,dashed] (1,0.9)-- (1,0.1);
\draw   [thick,->] (1.7,0.9)-- (1.7,0.1);
\draw   [thick,->] (2.4,0.9)-- (2.4,0.1);
\draw   [thick,->] (3.8,0.9)-- (3.8,0.1);
%%%%%%%%%%%%%%%%%%%%%%
\draw   [thick,->,dashed] (-2.5,1.9)-- (-1.85, 1.1);
\draw   [thick,->,dashed] (-1.8,1.9)-- (-1.15, 1.1);
\draw   [thick,->,dashed] (-1.1,1.9)-- (-0.7, 1.45);
\draw   [thick,->,dashed]  (-0.15,1.7)-- (0.25, 1.1);
\draw   [thick,->,dashed] (0.3,1.9)-- (0.95, 1.1);
\draw   [thick,->] (1,1.9)-- (1.65, 1.1);
\draw   [thick,->] (1.7,1.9)-- (2.35, 1.1);
%\draw   [thick,->] (1,1.9)-- (1.45, 1.1);
\draw   [thick,->](2.4,1.9)-- (2.7, 1.45);
\draw   [thick,->] (3.35,1.7)-- (3.75, 1.1);
%%%%%%%%%%%%%%%%%%%%%%%%%%
\node[draw =none] at (-0.4,1 ) {$\ldots$};
\node[draw =none] at (-0.4,2) {$\ldots$};
\node[draw =none] at (3,1 ) {$\ldots$};
\node[draw =none] at (3,2 ) {$\ldots$};
%\draw (t1)--(s1);
\end{tikzpicture}
%\vspace*{-5ex}

  \caption{Receiver conferencing scheme.}
  \label{fig1-1}
 % \vspace*{-2ex}
\end{figure*}~~%
\section{Proof of Theorem~\ref{lemma3}} \label{sec:rxconf}

The converse follows directly from Theorem~\ref{thm:converse} and \cite[Theorem 2]{Wiggeretal}. Achievability is proved in the following.

When transmitting only ``fast" messages or only ``slow" messages, the setup in this paper coincides with the setup in \cite{Wiggeretal} with 0 transmitter conferencing rounds and either 0 or  $\km$ receiver conferencing rounds.    Thus, by \cite{Wiggeretal},  the following two multiplexing gain pairs are achievable:
\begin{align}\label{eq:1}
&\Big(\S^{(F)}= \frac{1}{2}, \ \S^{(S)} = 0\Big),\\
&\Big(\S^{(F)}= 0, \ \S^{(S)} = \min\bigg\{  \frac{1}{2} + {\mu},\;\; \frac{ 2\km +1}{2\km+2} \bigg\} \Big).\label{eq:2}
\end{align}

Recall that in the case of only receiver conferencing, the coding scheme in \cite{Wiggeretal} periodically silences every  $2 \km+2$nd transmitter. This splits the network into smaller subnets of $2\km +1$ active transmitters and $2 \km+2$ receivers, where each active  transmitter can send a message at prelog 1. One of these subnets is depicted in Fig. \ref{fig1-1}. A close inspection of the coding scheme in \cite{Wiggeretal} reveals that the decoding of the  source messages sent at the left-most transmitter of each subnetwork does not rely on the conferencing messages. We can thus easily adopt the coding scheme in \cite{Wiggeretal} to our setup with ``fast" and ``slow" messages by letting the left-most transmitter of any subnetwork sends a ``fast" message and all other active transmitters send  ``slow" messages. {\color{black}To make the scheme more clear, we describe the communication in a given subnet at hand of  Fig. \ref{fig1-1}. Tx~$(1)$ encodes its ``fast'' message $M_1^{(F)}$ using a Gaussian codebook. It sends the resulting codeword $X_1^n(M_1^{(F)})$ over the channel. Each Tx~$(k), k \in\{2,\ldots,\km+1\}$, encodes its ``slow'' message $M_k^{(S)}$ using a Gaussian codebbok and then sends the resulting codeword $X_k^n(M_1^{(S)})$ over the channel. As we deactivated the last transmitter in the previous subnet, Rx~(1) observes the interference-free channel outputs $Y_1^{n} = X_1^n + Z_1^{n}$, based on which it decodes its desired message $M_1^{(F)}$.  After this decoding step, Rx~(1) sends its guess $\hat M_1^{(F)}$ to Rx~(2) during the first conferencing round. So,  $V_{1\rightarrow2}^{(1)} = \hat M_1^{(F)}$.  Rx~(2) uses this conferencing message to form $\hat Y_2^n = Y_2^n - \alpha X_1^n(\hat{M}_1^{(F)})$, based on which it  decodes its desired source message $M_2^{(S)}$.  The same procedure is applied at each Rx~$(k), k \in \{3,\ldots,\km+1\}$. Notice that Rx~$(k), k\in\{2,\ldots,\km+1\}$, sends its conferencing message $V_{k\rightarrow k+1}^{(k)} = \hat M_k^{(S)}$ in conferencing round $k$. 
\par We now explain the decoding of messages $M_{\km+2}^{(S)}, \ldots,M_{2\km+1}^{(S)}$. Recall that Tx~$(2\km+2)$ is silenced, and therefore $X_{2 \km+2}^n=0^n$. As a consequence, Rx~$(2\km+2)$ observes the channel outputs $Y_{2\km+2}^n = \alpha X^n_{2\km+1} + Z^n_ {2\km+2}$. Based on this outputs it decodes source message  $M_{2\km+1}^{(S)}$ and transmits it to Rx~$(2\km+1)$ over the conferencing link in round 1. So $V_{2\km+2\rightarrow 2\km+1}^{(1)} = \hat M_{2\km+2}^{(S)}$. Rx~$(2\km+1)$ uses the received conferencing message to form $\hat Y^n_{2\km+1} = Y^n_{2\km+1} -  X_{2\km+1}^n(\hat M_{2\km+1}^{(S)})$ and it decodes message $M_{2\km}^{(S)}$ based on this difference. It then sends the second-round conferencing message $V_{2\km+1 \to 2\km}^{(2)}=\hat{M}_{2 \km}^{(S)}$ to Rx~$(2\km)$. The same procedure is subsequently applied at Rxs $(2\km), (2\km-1), (2\km-2), \ldots,(\km+2)$. In particular,  each Rx~$(k), k\in \{2\km,\ldots,\km+2\}$, sends conferencing message $V_{k\rightarrow k-1}^{(k)} = \hat M_k^{(S)}$ in conferencing round $k$.}

%following three steps:
%\begin{enumerate}
%\item It reconstructs interference $\alpha_k X_{k-1}^n(\hat{M}_{k-1})$ using the conferencing message $\hat{M}_{k-1}$ obtained from Rx~$k-1$. 
%\item It forms  the presumingly interference-free signal
%\begin{equation*}
%\hat{Y}_{k}^n=Y_k^n-\alpha_{k} \hat{X}_{k-1}^n(\hat{M}_{k-1}),
%\end{equation*} and it decodes source message $M_k$  based on this difference.
%\item  It sends the decoded source message $\hat{M}_k$ over the conferencing link to Rx~$(k+1)$. The last receiver, \rt{Rx~$(\i+\kappaR)$}, does not send anything over the conferencing link (it skips this third step).
%\end{enumerate}  }.
With conferencing prelog  
%\begin{equation}
$\mu_{\max} = \frac{\km}{2\km+2}$,
%\end{equation} 
this scheme achieves the pair 
\begin{equation}\label{eq:3}
\Big(\S^{(F)}= \frac{ 1}{2 \km+2} , \ \S^{(S)} = \frac{2\km}{ 2\km+2}\Big).
\end{equation}
For a given conferencing prelog $\mu  \leq \mu_{\max}$, we timeshare this scheme with the scheme achieving \eqref{eq:1}. As a result,
%\begin{equation}
%\beta := \frac{\mu}{\mu_{\max}} = \mu\frac{ 2\km +2}{\km}
%\end{equation}   ensures that the conferencing prelog constraint \eqref{eq:conference_capa} is satisfied. We obtain that 
 for all  $\mu  \leq \mu_{\max}$, 
the following  pair of multiplexing gains is achievable:
\begin{subequations}\label{eq:3-2}
	\begin{IEEEeqnarray}{rCl}
		\S^{(F)} & := &  \beta \cdot \frac{1}{2\km +2} + (1-\beta) \cdot \frac{1}{2} =\frac{1}{2} - \mu\\[1.2ex]
		\S^{(S)} & := &  \beta \cdot \frac{2\km}{2\km +2} + (1-\beta)\cdot  0= 2\mu.
	\end{IEEEeqnarray}
\end{subequations}
Timesharing  finally the schemes achieving the pairs in \eqref{eq:1}, \eqref{eq:2} and \eqref{eq:3} establishes the direct part of the theorem. 

\begin{figure*}[t]
%\small
  \centering
  \small
    %\hspace*{32pt}
 \begin{tikzpicture}[scale=2.2, >=stealth]
\centering
\tikzstyle{every node}=[draw,shape=circle, node distance=0.5cm];
\draw [fill=gray!50](-2.5, 2) circle (0.1);
\draw [fill=gray!50](-1.8, 2) circle (0.1);
\draw [fill=gray!50](-1.1, 2) circle (0.1);
\draw [fill=gray!50](0.3, 2) circle (0.1);
\draw [fill=gray!50](1, 2) circle (0.1);
\draw [fill=gray!50](1.7, 2) circle (0.1);
\draw [fill=gray!50](2.4, 2) circle (0.1);
\draw [fill=white](3.8, 2) circle (0.1);
\node[draw =none, rotate =45] (s2) at (3.8,2) {\huge$+$};
%%%%%%%%%%%%%%
\draw [fill=black](-2.5, 0) circle (0.1);
\draw [fill=black](-1.8, 0) circle (0.1);
\draw [fill=black](-1.1, 0) circle (0.1);
\draw [fill=black](0.3, 0) circle (0.1);
\draw [fill=white](1,0) circle (0.1);
\node[draw =none, rotate =45] (s2) at (1,0) {\huge$+$};
\draw [fill=black](1.7,0) circle (0.1);
\draw [fill=black](2.4, 0) circle (0.1);
\draw [fill=black](3.8, 0) circle (0.1);
%%%%%%%%%%%%%%%%%%%
\node[draw =none] (s2) at (-2.5,1 ) {$+$};
\draw (-2.5, 1) circle (0.1);
\node[draw =none] (s2) at (-1.8,1 ) {$+$};
\draw (-1.8,1) circle (0.1);
\node[draw =none] (s3) at (-1.1,1 ) {$+$};
\draw (-1.1, 1) circle (0.1);
\node[draw =none] (s4) at (0.3,1 ) {$+$};
\draw (0.3, 1) circle (0.1);
\node[draw =none] (s5) at (1,1 ) {$+$};
\draw (1,1) circle (0.1);
\node[draw =none] (s6) at (1.7,1 ) {$+$};
\draw (1.7,1) circle (0.1);
\node[draw =none] (s7) at (2.4,1 ) {$+$};
\draw (2.4,1) circle (0.1);
\node[draw =none] (s8) at (3.8,1 ) {$+$};
\draw (3.8,1) circle (0.1);
%%%%%%%%%%%%%%%%%%%%%%%%%%%%%
\draw   [thick,->] (-2.4,2)-- (-1.9,2);
\draw   [thick,->] (-1.7,2)-- (-1.2,2);
\draw   [thick,->] (-1,2)-- (-0.5,2);
\draw   [thick,->] (-0.3,2)-- (0.2,2);
%\draw   [thick,->] (1.1,2)-- (-0.1,2);
\draw   [thick,->] (3.7,2)-- (3.2,2);
\draw   [thick,->] (3,2)-- (2.5,2);
\draw   [thick,->] (2.3,2)-- (1.8,2);
\draw   [thick,->] (1.6,2)-- (1.1,2);
%%%%%%%%%%%%%%%%%%%%%%%%%%%%%
\node[draw =none]  at (-2.4,2.22) {\small$M_1^{(S)}$};
\node[draw =none]  at (-1.7,2.22) {\small$M_2^{(S)}$};
\node[draw =none]  at (-1,2.22) {\small$M_3^{(S)}$};
\node[draw =none]  at (0.4,2.22) {\small$M_{\km+1}^{(F)}$};
\node[draw =none]  at (1.1,2.22) {\small$M_{\km+2}^{(S)}$};
\node[draw =none]  at (1.8,2.22) {\small$M_{\km+3}^{(S)}$};
\node[draw =none]  at (2.5,2.22) {\small$M_{\km+4}^{(S)}$};
%%%%%%%%%%%%%%%%%%%%%%%%%%%%
\node[draw =none]  at (-2.5,-0.25) {\small{1}};
\node[draw =none]  at (-1.8,-0.25) {\small2};
\node[draw =none]  at (-1.1,-0.25) {\small3};
\node[draw =none]  at (0.3,-0.25) {\small{$\km+1$}};
\node[draw =none]  at (3.8,-0.25) {\small{$2\km+2$}};
%%%%%%%%%%%%%%%%%%%%%%%%%%%%%%
\draw   [thick,->] (-2.5,1.9)-- (-2.5,1.1);
\draw   [thick,->] (-1.8,1.9)-- (-1.8,1.1);
\draw   [thick,->] (-1.1,1.9)-- (-1.1,1.1);
\draw   [thick,->] (0.3,1.9)-- (0.3,1.1);
\draw   [thick,->,dashed] (1,1.9)-- (1,1.1);
\draw   [thick,->,dashed] (1.7,1.9)-- (1.7,1.1);
\draw   [thick,->,dashed] (2.4,1.9)-- (2.4,1.1);
\draw   [thick,->,dashed] (3.8,1.9)-- (3.8,1.1);
%%%%%%%%%%%%%%%%%%%%%%%
\draw   [thick,->] (-2.5,0.9)-- (-2.5,0.1);
\draw   [thick,->] (-1.8,0.9)-- (-1.8,0.1);
\draw   [thick,->] (-1.1,0.9)-- (-1.1,0.1);
\draw   [thick,->] (0.3,0.9)-- (0.3,0.1);
\draw   [thick,->] (1,0.9)-- (1,0.1);
\draw   [thick,->] (1.7,0.9)-- (1.7,0.1);
\draw   [thick,->] (2.4,0.9)-- (2.4,0.1);
\draw   [thick,->] (3.8,0.9)-- (3.8,0.1);
%%%%%%%%%%%%%%%%%%%%%%
\draw   [thick,->,dashed] (-2.5,1.9)-- (-1.85, 1.1);
\draw   [thick,->,dashed] (-1.8,1.9)-- (-1.15, 1.1);
\draw   [thick,->,dashed] (-1,1.9)-- (-0.7, 1.45);
\draw   [thick,->,dashed] (-0.15,1.7)-- (0.25, 1.1);
\draw   [thick,->,dashed] (0.3,1.9)-- (0.95, 1.1);
\draw   [thick,->] (1,1.9)-- (1.65, 1.1);
\draw   [thick,->] (1.7,1.9)-- (2.35, 1.1);
%\draw   [thick,->] (2.4,1.9)-- (2.45, 1.1);
\draw   [thick,->] (2.4,1.9)-- (2.7, 1.45);
\draw   [thick,->] (3.35,1.7)-- (3.75, 1.1);
%%%%%%%%%%%%%%%%%%%%%%%%%%
\node[draw =none] at (-0.4,1 ) {$\ldots$};
\node[draw =none] at (-0.4,0) {$\ldots$};
\node[draw =none] at (3,1 ) {$\ldots$};
\node[draw =none] at (3,0 ) {$\ldots$};
%\draw (t1)--(s1);
\end{tikzpicture}
%\vspace*{-5ex}

  \caption{Transmitter conferencing scheme.}
  \label{fig1-2}
 % \vspace*{-2ex}
\end{figure*}
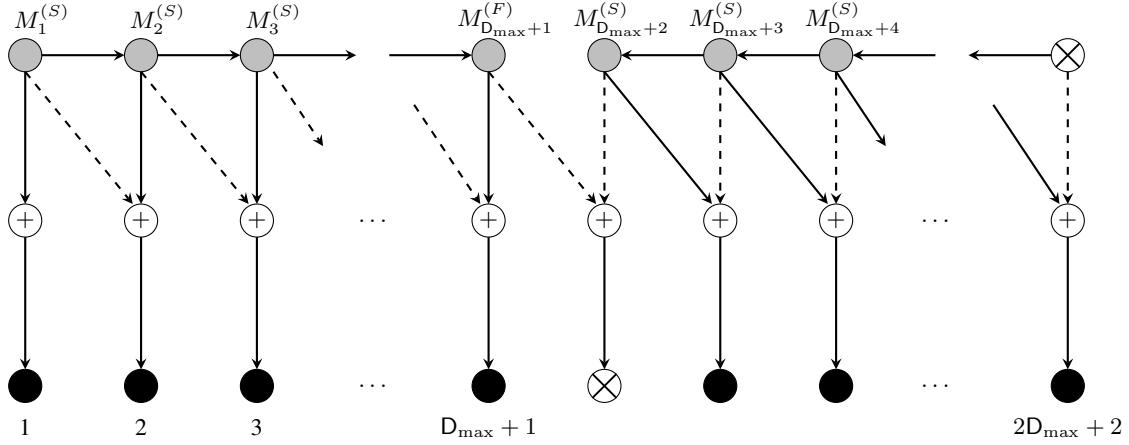~~%
{\color{black}
\section{Proof of Theorem \ref{thm:txconf}} \label{sec:txconf}
\subsection{Direct Part}
The proof is similar to the the proof in Section \ref{sec:rxconf}. The main difference is in the scheme achieving \eqref{eq:3}. In the following, we describe a scheme that achieves \eqref{eq:3} with transmitter conferencing but no receiver conferencing. As before, we silence every $2\km+2$nd transmitters, which splits the network into non-interfering subnetworks. In a given subnetwork we apply the scheme depicted in Fig. \ref{fig1-2}. Specifically, Tx~(1) encodes its ``slow'' message $M_1^{(S)}$ using a Gaussian point-to-point codebook, and sends the resulting codeword $X_1^n(M_1^{(S)})$ over the channel. It also quantises $X_1^n(M_1^{(S)})$ using a rate $1/2\log(1 +P)$ quantiser and sends the resulting quantisation message as a first-round conferencing message to Tx~(2). Upon receiving this quantisation message, Tx~(2) reconstructs the quantised input $\hat X_1^n(M_1^{(S)})$ and encodes $M_2^{(S)}$ using a power $P$ dirty-paper code that eliminates the interference $\alpha \hat X_1^n(M_1^{(S)})$. It then  sends this dirty-paper sequence over the channel. Moreover, Tx~(2) also  quantises its  input sequence $X_2^n$ using a rate $1/2\log(1 +P)$ quantiser, and sends the quantization message as a second-round conferencing message  to Tx~$(3)$. The procedure is repeated subsequently  for each Tx~$(k), k \in\{3,\ldots,\km\}$.  Tx~$(\km + 1)$ produces its inputs in a similar way, i.e., using dirty-paper coding to mitigate the interference $\alpha \hat X_{\km}^n$. But in contrast to the previous transmitters, it sends a ``fast'' message $M_{\km+1}^{(F)}$. Receivers~$1,2,\ldots, \km+1$ decode their intended messages using an optimal  dirty-paper decoding rule.

We now explain transmission of messages $M_{\km+3}^{(S)}, \ldots,M_{2\km+2}^{(S)}$. On a high level, Message $M_k^{(S)}$ is sent over the path Tx~$(k)$--Tx~$(k-1)$--Rx~$(k)$. This means, that the actual communication is performed over the ``interference links", whereas direct links carry interference. To indicate this, in Figure~\ref{fig1-2}, the ``interference" links are depicted in solid lines and the direct links in dashed lines.  More specifically, Tx~$(2\km+2)$  encodes its ``slow'' message $M_{2\km+2}^{(S)}$ using a Gaussian point-to-point codebook, and it sends a quantisation message describing the codeword $X_{2\km+1}^n(M_{2\km+1}^{(S)})$ as a first-round conferencing message to Tx~$(2\km+1)$.  Tx~$(2\km+2)$ does not send any channel inputs, i.e., $X_{2\km+2}^n=0^n$. Tx~$(2\km+1)$ sends the quantised sequence that corresponds to the conferencing message it received from Tx~$(2\km+2)$ over the channel. It then encodes its own ``slow" message $M_{2\km+1}^{(S)}$ using a dirty-paper code that cancels its own transmit signal $X_{ 2 \km+1}^n$ as interference. Finally, Tx~$(2\km+1)$ quantises this produced dirty-paper sequence and sends the quantisation message as a second-round conferencing message to Tx~$(2\km)$. This latter sends the quantisation sequence that corresponds to its received conferencing message as inputs $X_{2\km}^n$ over the channel. Transmission of messages $M_{2\km}^{(S)}, \ldots, M_{\km+2}^{(S)}$ is performed in a similar way. All receivers decode their intended messages using an optimal dirty-paper decoding rule.

%
%In this scheme  Tx~$k \in\{1,\ldots,\km\}$ performs the following three steps:
%\begin{enumerate}
%\item Using the conferencing message from Tx~$(k-1)$, it reconstructs the quantised signal $\hat{M}^{(S)}_{k-1}$. 
%
%\item It encodes and transmits its source message  $X_k^n$ using a power-$P$ dirty-paper code that eliminates  interference $\alpha \hat{M}_{k-1}^{(S)}$.  
%\item It quantises its ``slow'' message $M_k^{(S)}$ with a rate $(1/2) \log (1+P)$ quantiser and sends it over the conferencing link  to transmitter $k+1$. 
%\end{enumerate} 
% Transmitter $(\km + 1)$  sends ``fast'' message but this message is never used in the conferencing messages. 
% 
% Each transmitter $k \in\{\km+2,\ldots, 2\km+1\}$  performs following steps:
% \begin{enumerate}
%\item Using the conferencing message from transmitter $(k+1)$, it reconstructs the quantised signal $\hat{M}_{k+1}^{(S)}$. 
%\item It transmits $X_k^n$ over the network.
%\item  It encodes message $M_k^{(S)}$ using dirty-paper coding and sends the resulting quantisation bits over the conferencing link to Transmitter $(k-1)$.
%\end{enumerate}  
%Finally, each receiver decodes its desired source message based on its channel output. 

\subsection{Converse}
Bound \eqref{a2} follows directly from \cite{Wiggeretal}, which showed that the bound holds even when only ``slow'' messages are transmitted. In fact, the sum of the rates of ``slow'' and ``fast'' messages cannot exceed the largest achievable rate of ``slow'' messages only. 

We are left with proving that inequality \eqref{a1} remains valid with transmitter conferencing. Let $\mathbf{M}^{(S)} := (M_1^{(S)},\ldots,$ $M_K^{(S)})$. By Fano's Inequality and the independence of the messages, for any $k \in \{1,\ldots, K-1\}$:
\begin{IEEEeqnarray}{rCl}\label{eq:conv2}
\lefteqn{R_k^{(F)} + R_{k+1}^{(S)} +  R_{k+1}^{(F)} }\nonumber\\
 &= & \frac{1}{n} \Big[H(M_k^{(F)} | \mathbf{M}^{(S)}, M_{k-1}^{(F)})\nonumber\\ & & \qquad + H(M_{k+1}| M_1^{(S)},  \ldots,M_{k}^{(S)}, M_{k+2}^{(S)},\ldots, M_K^{(S)})\Big]  \nonumber \\& \leq& \frac{1}{n} \Big[ I(M_k^{(F)}; Y_k^{n} |  \mathbf{M}^{(S)},M_{k - 1}^{(F)})    \nonumber \\  & &  \qquad + I(M_{k+1}; Y_{k+1}^{n}  |M_1^{(S)}, \ldots, M_{k}^{(S)}, M_{k+2}^{(S)},\ldots, M_K^{(S)}) \Big] \nonumber\\ & & \qquad \frac{\epsilon_n}{n}\nonumber \\ &=& \frac{1}{n} \Big[ h(X_k^n + Z_k^n| \mathbf{M}^{(S)}) - h(Z_k^n)  \nonumber \\ & & \qquad +h(Y_{k+1}^n| M_1^{(S)}, \ldots, M_{k}^{(S)}, M_{k+2}^{(S)},\ldots, M_K^{(S)})  \nonumber \\
 & & \qquad - h (\alpha X_k^n + Z_{k+1}^n | \mathbf{M}^{(S)})\Big]\\ & & + \frac{\epsilon_n}{n}\nonumber \nonumber \\ 
& \stackrel{(a)}{\leq } &  \frac{1}{2} \log (1+ (1 + |\alpha|)^2 P) \nonumber \\ 
&  &+ \text{max} \{ - \log |\alpha|, \log(\alpha^2 -1) +  \frac{\epsilon_n}{n},
\end{IEEEeqnarray}
where $(a)$ follows by the entropy-maximizing property of the Gaussian distribution and by \eqref{ee2} and \eqref{ee4}.
One can also prove that
\begin{equation} \label{Kf}
R_K^{(F)} + R_{K}^{(S)} \le \frac{1}{2} \log (1+ (1 + |\alpha|^2) P) + \frac{\epsilon_n}{n}.
\end{equation}
 We now sum up the bound in \eqref{eq:conv2} for all values of $k \in \{1,\ldots, K-1\}$ and combine it with \eqref{Kf}.  Taking  $n \to \infty$,  it follows  that whenever the probability of error $P_e^{(n)}$ vanishes as $n\to \infty$ (and thus $\frac{\epsilon_n}{n}\to 0$ as $n\to \infty$):
\begin{IEEEeqnarray}{rCl} \label{bn2}
\lefteqn{\sum_{k = 1}^K \left(2 R_k^{(F)} + R_k^{(S)}\right) }\nonumber \\
& = & R_1^{(F)} + \sum_{k=1}^{K-1} \Big(R_k^{(F)}+ R_k^{(S)}+ R_{k+1}^{(F)}\Big) + R_K^{(F)} + R_K^{(S)}    \notag\\
& \leq &  (K-1)\frac{1}{2} \log (1+(1+ \alpha^2) P)  + \log( 1+P)  \nonumber\\
&+ & (K-1)  \max \{ -  \log |\alpha| , 0\},\IEEEeqnarraynumspace
\end{IEEEeqnarray} 
Dividing by $K$ and taking $P,K \to \infty$, the converse to \eqref{a1} is established.}
\section{Conclusion}We presented upper and lower bounds on the capacity region of Wyner's soft-handoff network with receiver conferencing under mixed decoding constraints. Our results show that when the messages with the stringent decoding delay have small or moderate rates, then there is no penalty in sum-rate caused by this stringent decoding delay. When the rate of these messages is large, then any rate increase $\Delta$ requires that the rate of the messages with non-stringent delay constraint be reduced by approximately $2\cdot \Delta$. 
The paper also characterizes the  multiplexing gain region with mixed decoding delays with either transmitter conferencing or receiver conferencing. The two multiplexing gain regions for the two setups coincide and thus exhibit  a  duality between transmitter and receiver conferencing in this mixed-delay setup. 
%
%\section{Outlook}
Extending these results to a setup with both transmitter and receiver conferencing seems interesting. In particular, our preliminary results indicate that the gains with both types of conferencing are much more pronounced than with only transmitter or only receiver conferencing. Particular interest will also be on duality-aspects of transmitter and receiver conferencing with respect to the multiplexing gain.

\section*{Acknowledgement}
The work of H. Nikbakth and M. Wigger was supported by the ERC grant CTO Com. The work of S. Shamai has been supported by the European Union's Horizon 2020 Research And Innovation Programme, grant agreement no. 694630.

\end{document}